\documentclass[fleqn,usenatbib]{mnras}
\usepackage{savesym}
\usepackage{amsmath}
\savesymbol{iint}
\savesymbol{iiint}
\usepackage{txfonts}
\restoresymbol{TXF}{iint}
\restoresymbol{TXF}{iiint}

\usepackage[T1]{fontenc}
\usepackage{ae,aecompl}

\usepackage{graphicx}	

\def\lsim{\mathrel{\rlap{\lower 4pt \hbox{\hskip 1pt $\sim$}}\raise 1pt \hbox
        {$<$}}}
\def\gsim{\mathrel{\rlap{\lower 4pt \hbox{\hskip 1pt $\sim$}}\raise 1pt \hbox
        {$>$}}}

\newcommand{\eg}{e.g.,\ }

\newcommand{\Msun}{M_{\odot}}

\newcommand{\Ha}{H$\alpha$}

\newcommand{\HII}{H~{\sc ii}}

\newcommand{\CII}{C~{\sc ii}}

\newcommand{\SiII}{Si~{\sc ii}}

\newcommand{\CaII}{Ca~{\sc ii}}

\newcommand{\FeII}{Fe~{\sc ii}}
\newcommand{\FeIII}{Fe~{\sc iii}}

\newcommand{\NiII}{Ni~{\sc ii}}

\newcommand{\Nifs}{$^{56}$Ni}

\newcommand{\KE}{$E_{\rm kin}$}

\title[SN\,1986G]{Abundance stratification in Type Ia supernovae - V. SN\,1986G bridging the gap between normal and subluminous SNe\,Ia}
\author[C. Ashall]{C.Ashall$^{1}$\thanks{E-mail:c.ashall@2013.ljmu.ac.uk}, P.A. Mazzali$^{1,2}$, E.Pian$^{3,4}$, P.A.James$^{1}$\\
$^{1}$Astrophysics Research Institute, Liverpool John Moores University, IC2, Liverpool Science Park, 146 Brownlow Hill, \\  Liverpool L3 5RF, UK\\
$^{2}$Max-Planck-Institut f\"ur Astrophysik, Karl-Schwarzschild-Str. 1, D-85748 Garching, Germany\\
$^{3}$Institute of Space Astrophysics and Cosmic Physics, via P. Gobetti 101, I-40129 Bologna, Italy\\
$^{4}$Scuola Normale Superiore, Piazza dei Cavalieri 7, I-56126 Pisa, Italy\\ }

\begin{document}

\date{July 2016}

\pagerange{\pageref{firstpage}--\pageref{lastpage}} \pubyear{2016}

\maketitle

\label{firstpage}

\begin{abstract} A detailed spectroscopic analysis of SN\,1986G has been
performed. SN\,1986G `bridges the gap' between normal and sub luminous type Ia
supernova (SNe\,Ia). The abundance tomography technique is used to determine the
abundance distribution of the elements in the ejecta. SN\,1986G was found to be
a low energy Chandrasekhar mass explosion. Its kinetic energy was 70\% of the
standard W7 model ($0.9\times$10$^{51}$\,erg).  Oxygen dominates the ejecta from
the outermost layers down  to $\sim$ 9000\,kms$^{-1}$ , intermediate mass
elements (IME) dominate from $\sim$ 9000\,kms$^{-1}$ to  $\sim$ 3500\,kms$^{-1}$
with Ni and Fe dominating the inner layers $<\sim$ 3500\,kms$^{-1}$.  The final
masses of the main elements in the ejecta were found to be, O=0.33\,M$_{\sun}$,
IME=0.69\,M$_{\sun}$, stable NSE=0.21\,M$_{\sun}$, $^{56}$Ni=0.14\,M$_{\sun}$.
An upper limit of the carbon mass is set at C=0.02\,M$_{\sun}$. The spectra of
SN\,1986G consist of almost exclusively singly ionised species. SN\,1986G can be thought of as a low luminosity
extension of the main population of SN\,Ia, with a large deflagration phase 
that produced more IMEs than a standard SN\,Ia.   
\end{abstract}


\section{Introduction}

Type Ia Supernovae (SNe\,Ia) are predicted to originate from the thermonuclear
explosion of a C+O white dwarf (WD)  approaching the Chandrasekhar mass limit,
but their exact nature remains unclear. There are currently two leading 
progenitor scenarios, the single degenerate scenario (SDS) and the double
degenerate scenario (DDS).  In the SDS a C+O WD accretes material from a
non-electron-degnerate companion star. When the mass of the C+O WD approaches
the Chandrasekhar mass thermonuclear runaway begins, hence a SN Ia is produced
\citep{Nomoto97}. In the DDS two WDs merge, after losing angular momentum in the
form of gravitational waves, and explode \citep{WDWD}.  Additional scenarios
include the Sub-Chandrasekhar mass explosion of a C+O WD which accretes Helium
from a companion \citep{livne95}, and the collision of two C+O WDs in a triple
system, which can detonate following a head-on collision \citep{Rosswog09}.  

Historically it was thought that SNe\,Ia could be used as standard candles.
However, as with much of observational astronomy, when the sample of data
increased it soon became apparent that they are a far more diverse group
\citep{phillips87,Phillips92,1991bg,Ashall16}. Normal SNe\,Ia show a strong
correlation between their light curve shape, $\Delta$M$_{15}$($B$), and absolute
magnitude. This correlation is one of the underlying foundations of SN\,Ia
cosmology \citep{phillips93,phillips1999}. There are, however, a number of 
subsets of SNe\,Ia which are less common. Among these, SN\,1991bg, after which a
subset of SNe\,Ia are named, was a less luminous explosion which was dominated
by intermediate mass elements and showed a rapidly declining light curve.
Theoretically SN\,1991bg has been interpreted to be the merger of two WDs
\citep{Mazzali12}, but others have suggested a delayed detonation mechanism
\citep{Hoflich02}.

SN\,1986G was the first object to lead to the questioning of Type Ia Supernovae
(SNe\,Ia) as standard candles \citep{phillips87}. Its light curve (LC) was 
faster and dimmer than all of the previously discovered SNe\,Ia. It had a
rapidly declining light curve at early phases and slow expansion velocities
compared to a normal SNe\,Ia \citep{phillips87,Cristiani92}. SN\,1986G was
located in NGC 5128, also known as Centauras A, which is at a distance
3.42$\pm$0.18\,Mpc \citep{Ferrarese07}. This made it one of the closest SNe\,Ia ever discovered,
until SN 2014J \citep{Ashall14}.  SN\,1986G was observationally red, with a
$(B-V)_{Bmax}$=0.88$\pm$0.03\,mag. This could be interpreted as SN\,1986G 
suffering form a large amount of extinction. However, SNe\,Ia suffer from a
colour/extinction degeneracy. Recently, for normal SNe\,Ia, this
colour/extinction degeneracy has been overcome \citep{Sasdelli16}. For unusual
SNe\,Ia this degeneracy still remains an issue, and values of host galaxy
extinction can be very uncertain.   

\begin{figure}
\centering
\includegraphics[scale=0.45]{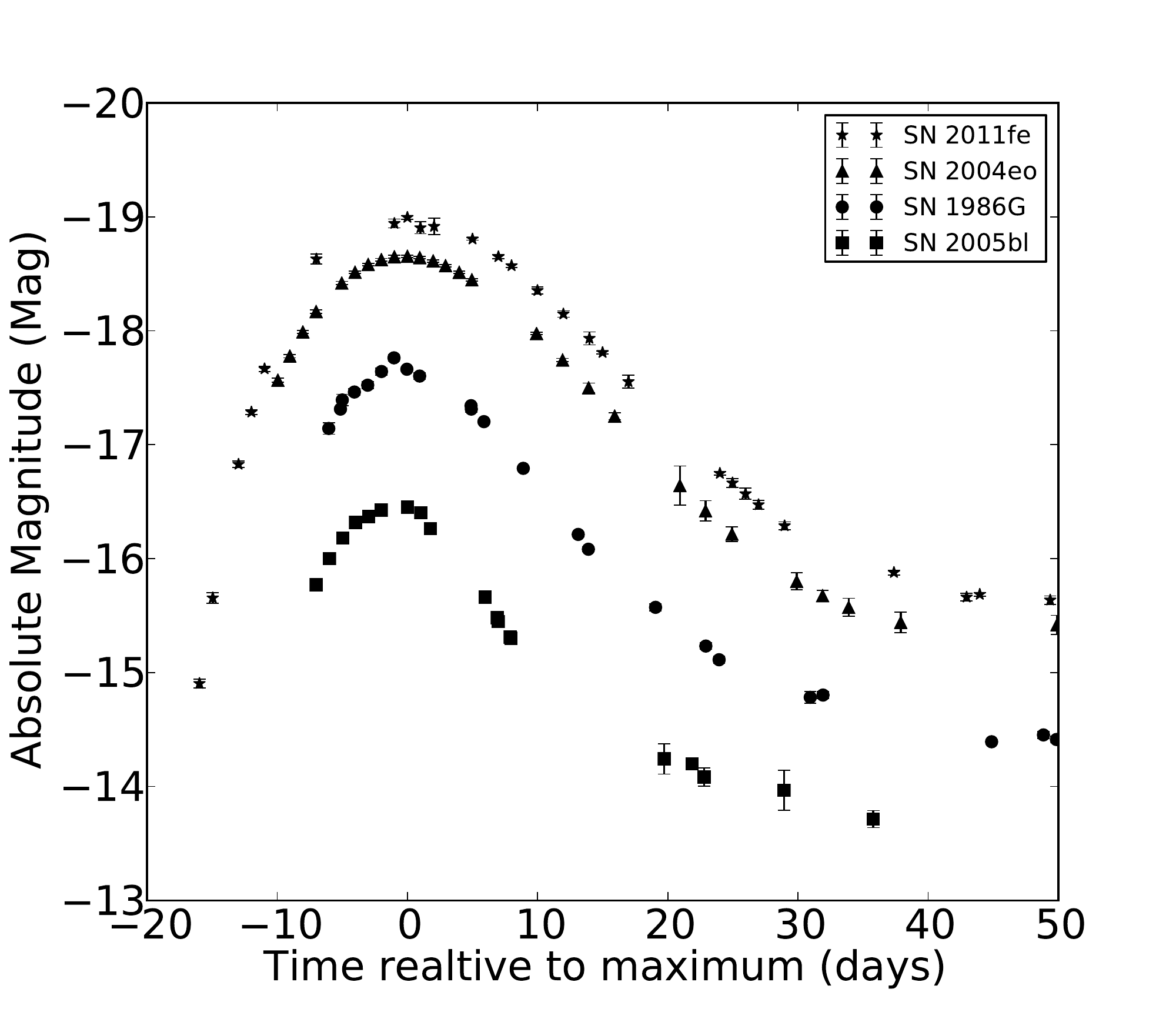}
\caption{The $B$ band absolute magnitude of four SNe\,Ia which have a variety of LC shapes.}
\label{fig:photcomp}
\end{figure}

\begin{figure}
\centering
\includegraphics[scale=0.45]{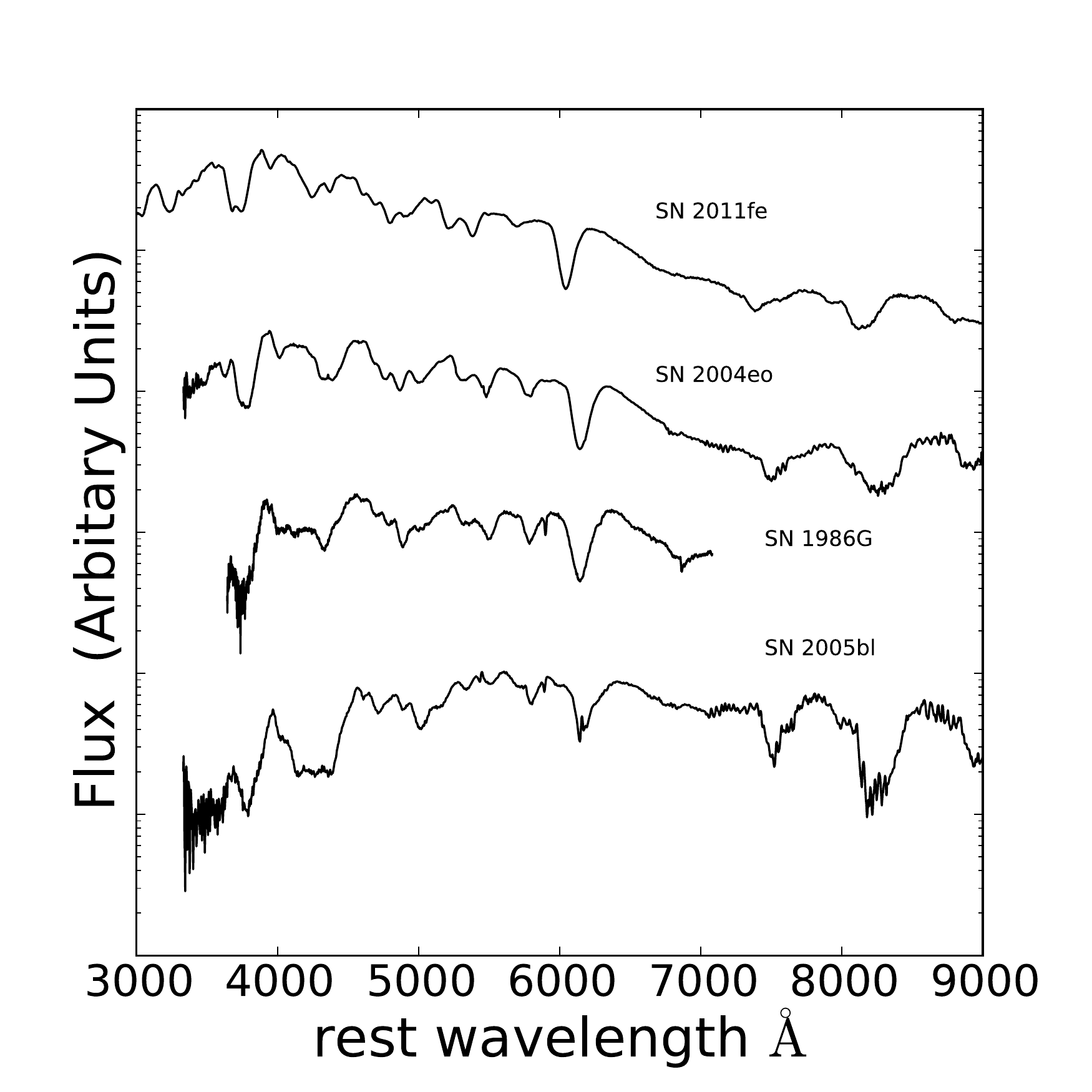}
\caption{The spectra of four SNe\,Ia at $B$ band maximum. The SNe correspond to the 4 SNe in Figure \ref{fig:photcomp}}
\label{fig:spectracomp}
\end{figure}

SN\,1986G is a ``transitional'' SN\,Ia. ``transitional'' in this case refers to
SNe\,Ia that bridge the gap between 91bg-like and normal SNe\,Ia
\citep{Ashall16}.  Figure \ref{fig:photcomp} presents the $B$-band absolute
magnitude light curves of four SNe\,Ia with a variety of decline rates,
$\Delta$M$_{15}$($B$) (2011fe, 2004eo, 1986G and 2005bl).  SN 2011fe was a
stereotypical normal SNe\,Ia with $\Delta$M$_{15}$($B$)=1.1$\pm$0.05\,mag, a
broad LC shape, and a normal spectrum, see Figure \ref{fig:spectracomp}.
SN\,2004eo had $\Delta$M$_{15}$($B$)=1.47$\pm$0.07\,mag, it was slightly less
luminous than SN\,2011fe, and had normal spectra. SN\,1986G had 
$\Delta$M$_{15}(B)$=1.81$\pm$0.07\,mag, a narrow LC and a cooler spectrum.
SN\,2005bl was a 91bg-like SN, with $\Delta$M$_{15}(B)$=1.93$\pm$0.1\,mag.
SN\,1986G sits half-way between normal and subluminous SNe\,Ia.

The environment of SN\,1986G is highly unusual and merits some discussion. 
The host system is the peculiar radio galaxy NGC\,5128 
which is generally classified in terms of its optical morphology as a lenticular or 
elliptical galaxy.  This implies a dominant old stellar population, but this may well be
misleading, as SN\,1986G occurred in the middle of the strong dust lane which gives
NGC\,5128 its peculiar appearance, but well offset from the active galactic nucleus. 
The dust lane is generally considered to be the result of a merger with a smaller 
gas-rich galaxy \citep{Baade54, Tubbs80}, 
and it is associated with substantial ongoing star formation, which is indicated 
by strong, extended clumpy H$\alpha$ emission throughout most of the dust lane 
\citep[see \eg][]{Bland87}.  \citet{Anderson15}
found the location of SN\,1986G  to be associated with detectable H$\alpha$ emission,
even though it is not coincident with a bright \HII\ region, so there is clearly a young stellar
population in the immediate vicinity.  Transitional SNe\,Ia are often found in peculiar galaxy
environments, and this is clearly the case with SN\,1986G.

Figure \ref{fig:spectracomp} shows the spectroscopic difference between these
SNe\,Ia. All spectra have been corrected for host galaxy extinction.  A strong
$\sim$4450\,\AA\ Ti II feature is an indicator of a subluminous SN\,Ia.
SN\,1986G is one of the few SNe\,Ia with an intermediate strength Ti II feature.
Furthermore, the ratio of the Si II features ($\sim$5970\,\AA\ and 6355\,\AA\ )
is a good temperature indicator. \citep{Nugent05,hanchlingerEW}. It is however
an indirect temperature indicator, as it results from the saturation of the Si
II $\lambda$ 6355 line.  Si II becomes more abundant with respect to Si III at lower
luminosities and temperatures, so that the Si II $\lambda$5970 line becomes stronger with
respect to the Si $\lambda$6355 line.  SN\,1986G had a larger Si II ratio compared
to normal SNe\,Ia, but one not as strong as a 91bg-like SN\,Ia. Therefore it is
a transitional object. SN\,1986G is the only published SNe\,Ia with these
properties and good observational data, which demonstrates that these objects
must be rare. SN\,2011iv may also be classified as a transitional SN because of
its rapidly evolving LC, but it was spectroscopically normal \citep{Foley12}.  

A detailed theoretical understanding of SN\,1986G should help to determine the
relationship between 91bg-like and normal SNe\,Ia. This can be done by examining
a time series of spectra and analysing their evolution.  Normal SNe\,Ia show
properties that are matched by delayed detonation (DD) explosions, and can be
modelled using the W7 density profile, which is similar to most DD models except
in the outermost layers  \citep{Stehle2005,Mazzali08,Ashall14}. 
SN\,1991bg had a low central density \citep{Mazzali12}, which favours a merger
scenario in which the combined mass of the merging WDs is below the
Chandrasekhar mass. Similarly, the density distribution in SN\,2003hv, which
had  $\Delta$M$_{15}$($B$)=1.61\,mag, indicates a  Sub-Chandrasekhar mass
\citep{Mazzali11}. Subluminous SNe\,Ia tend to occur in old stellar populations
and are at the end of the SNe\,Ia parameter space \citep{Ashall16}. Therefore,
as it has been shown that 91-bg-like SNe\,Ia could be the result of a different
progenitors/explosion mechanism than normal SNe\,Ia, the question is at which
point do SNe\,Ia begin to depart from the standard explosion models, and what
causes them to be different. The answer may be found by analysing individual
transitional SNe\,Ia in detail.  One way to do this is to use the `abundance
tomography' approach \citep{Stehle2005}. This method produces optimally fitting
synthetic spectra to match the observed ones, and allows us to infer the
abundance distribution in the ejecta. Abundance tomography has been successfully
used to model a number of SNe\,Ia, including SN\,2002bo \citep{Stehle2005}, 
2003du \citep{2003du}, 2010jn \citep{hachinger2013}, 2011fe
\citep{mazzali2013,Mazzali15}, 2014J \citep{Ashall14} and 1991T
\citep{sasdelli14}.  

This paper uses the abundance tomography technique to infer the properties of
SN\,1986G. It starts with a summary of the data used (Section 2), followed by a
description of the modelling technique (Section 3). Photospheric-phase models
are then presented in Section 4, while Section 5 discusses nebular-phase models.
After this the full abundance tomography is carried out (Section 6).  In Section
7 the results are re-evaluates and a modified density profile is used to improve
the fits. A synthetic light curve is presented in Section 8 and compared to the 
bolometric light curve of SN\,1986G computed from the available photometry.
Finally, the results are discussed and conclusions are drawn in Section 9.

\section[data]{data}

\begin{table*}
 \centering
 \caption{The $\Delta$ M$_{15}$($B$) and absolute $B$ band magnitude of the SNe\,Ia used in Figures  \ref{fig:photcomp} and  \ref{fig:spectracomp}.}
  \begin{tabular}{cccc}
  \hline
   SN&$\Delta$M$_{15}(B)$&M$_{B}$&References\\
  \hline
 SN 2011fe&1.1$\pm$0.05\,mag&-18.99&\citet{munari2013}\\
 SN 2004eo&1.47$\pm$0.07\,mag&-18.65&\citet{Pastorello07}\\
 SN\,1986G&1.81$\pm$0.07\,mag&-17.76&\citet{phillips87,Taubenberger08}\\
 SN 2005bl&1.93$\pm$0.1\,mag&-16.45&\citet{Taubenberger08}\\
\hline
\end{tabular}

\label{table:LCinfo}
\end{table*}

Abundance tomography modelling requires a time-series of spectra. The data in
this paper come from a variety of sources, which are listed in Table
\ref{table:data}. Six spectra published by \citet{Cristiani92} were used for the
photospheric phase models. These spectra cover the range from $-3$\,d to $+2$\,d
relative to $B$-band maximum. This is not ideal as it only covers a small range
in velocity space, but information about the outer (but not the outermost)
layers can still be inferred.  Two near-UV spectra taken with the International
Ultraviolet Explorer (IUE) are also used. Although these spectra have low
signal-to-noise, they are important in that they allow us to determine the flux
level in the NUV. To probe the inner layers of the SN ejecta a nebular phase
spectrum is required. One such spectrum was obtained by \citet{Cristiani92}. The
final stage of the modelling process involves modelling the bolometric light
curve. The data used to construct the pseudo-bolometric light curve of SN\,1986G
are taken from a variety of sources \citep{Cristiani92, Frogel87, phillips87}.

Before the spectra can be modelled they must be calibrated in flux. The flux
calibration of the spectra was checked by comparing synthetic photometry,
obtained from the spectra with the real photometric data. As SN\,1986G only has
good photometric coverage in two passbands, $B$ and $V$, flux calibration was
checked against the photometry in these filters. If there was a difference in
magnitude between the synthetic and observed photometry a linear function was
applied to correct the spectra. New synthetic $B$ and $V$ band photometry was
then produced and checked against the observed photometry. This process was
carried out for all spectra of SN\,1986G.

\begin{table*}
 \centering
 \caption{The spectra of SN\,1986G.}
  \begin{tabular}{ccccc}
  \hline
   Date&MJD&Phase (days)&source&telescope\\
   \hline
7/5/1986&46557&-3&\citet{Cristiani92}&ESO 2.2m B\&C+CCD\\
7/5/1986&46557&-3&archive.stsci.edu/iue/&International Ultraviolet Explorer\\
8/5/1986&46558&-2&\citet{Cristiani92}&ESO 2.2m B\&C+CCD\\
9/5/1986&46558&-2&archive.stsci.edu/iue/&International Ultraviolet Explorer\\
9/5/1986& 46559&-1&\citet{Cristiani92}&ESO 1.5m B\&C+IDS\\
10/5/1986&46560&+0&\citet{Cristiani92}&ESO 1.5m B\&C+IDS\\
11/5/1986&46561&+1&\citet{Cristiani92}&ESO 1.5m B\&C+IDS\\
12/5/1986&46562&+2&\citet{Cristiani92}&ESO 1.5m B\&C+IDS\\
22/1/1987&46817&+256&\citet{Cristiani92}&EFOSC 3.6m\\
\hline
\end{tabular}
\label{table:data}
\end{table*}

\section{Modelling Techniques}
\subsection{Photospheric phase modelling technique}

With the objective of carrying out a detailed analysis of the ejecta of
SN\,1986G, a MC supernova radiative transport code was used to produce synthetic
spectra, implementing the `abundance tomography' method  outlined in
\citet{Stehle2005}. This technique utilises the fact that as time from the SN
explosion increases and the SN ejecta expand, deeper and deeper layers can be
seen. The ejecta are assume to be in homologous expansion, which can be
approximated by the equation $r=v_{ph}\times t_{exp}$, where $r$ is the 
distance from the centre of the explosion, $v_{ph}$ is the photospheric velocity
and $t_{exp}$ is the time from explosion. The code uses the
Schuster-Schwarzschild approximation, which assumes that the radiative energy is
emitted from an inner boundary as a black body. This approximation is used as it
does not require in-depth knowledge of the radiation transport below the
optically thick photosphere while still yielding accurate results. The
assumption works best when the bulk of the $^{56}$Ni is below the photosphere,
since $^{56}$Ni decay produces the energy which powers the ejecta. At later 
times significant $\gamma$-ray trapping occurs above the photosphere. After
bolometric maximum the Schuster-Schwarzschild approximation can cause excess
flux in the IR, but since most of the strong lines in a SN\,Ia are in the
UV/Optical ($<6500\,\AA$) this does not affect the results concerning
abundances.

The code is a 1D MC radiative transport code
\citep{Abbott1984,mazzalilucy93,Lucy1998,mazzali2000}. It simulates the emission
of photon packets at the photosphere. These packets then propagate through the
SN atmosphere, where they can undergo Thomson scattering and line absorption. If
a packet is absorbed in a line it is re-emitted following a photon branching
scheme, which allows both fluorescence (blue to red) and reverse fluorescence
(red-to-blue) to occur. This is the process that mostly determined the spectrum
of a SN\,Ia \citep{mazzali2000}. Electron scattering is a secondary source of
opacity in the metal-rich atmosphere of SNe\,Ia \citep{Pauldrach96}. Packets
which go back into the photosphere are assumed to be re-absorbed.  A modified
nebular approximation is used to treat the excitation/ionisation state of the
gas, to account for NLTE effects caused by the diluted radiation field. The
radiation field and the state of the gas are iterated until convergence is
achieved. The final spectrum is obtained by computing the formal integral of the
radiation field.  

The code requires a number of free input parameters, as well as a fixed density
profile. A suitable density profile for the explosion must first be chosen.
Usually for a SNe\,Ia the initial density profile used is the W7 profile
\citep{w7}, which has been shown to produce good results for a range of
SNe\,Ia. The W7 density profile assumes a single-degenerate SN\,Ia explosion
\citep{w7}. If  W7 does not produce a good fit different density profiles are
then tested, such as a Sub-Chandrasekhar (Sub-Ch) mass density profile.  The
normal process of modelling a time series of spectra involves producing
synthetic spectra starting with the earliest epoch.  This is done by setting the
distance and extinction to the SN, then determining the bolometric luminosity of
the ejecta and the photospheric velocity. A typical abundance distribution is
then assumed and a synthetic spectrum of the first epoch is produced. Usually 2
or 3 synthetic shells are placed above the photosphere of the first observed
spectrum, in order to produce a stratified abundance distribution at higher
velocities, and to avoid wrong elements been seen at high velocities in the
synthetic spectra. The abundances above the photosphere are varied to produce
the best fit to the observed spectrum. Once an acceptable fit is obtained the
next spectrum is modelled using the same procedure, and only the abundances
inside the previous photosphere are modified. Abundances in the outer layers can
affect the synthetic spectra at later epochs. If this is the case iteration is
required on the abundances of the elements in the outer layers. Abundances given in
this work have an error of $\pm$25\%, and photospheric velocities have an error of $\pm$15\%. A detail
analysis of errors using the 'abundance tomography' method was carried out in the second paper of this series, see \citet{Mazzali08}.
The error on the \Nifs\ mass is $\pm$10\%, as it is constrained though multiple methods.
 For SN\,1986G this was done for all of the photospheric phase spectra, up to 2 days past
$B$-band maximum.

\subsection{Nebular Phase modelling technique }

The nebular spectrum was modelled using a 1D NLTE code similar to
\citet{mazzalineb}. The code uses a Monte Carlo scheme to compute the
propagation and deposition of the $\gamma$-rays and positrons production in the
decay of $^{56}$Ni to $^{56}$Co and $^{56}$Fe, based on an assumed density
profile, such as W7, and an abundance distribution. In the outer layers the
abundances derived in the photospheric phase modelling are used, while inner
layers are filled with the abundances that lead to a best match with the
observations \citep{mazzalineb}. The heating from the energy deposition is
balanced by the cooling from line emission. Given a fixed distance and
extinction, the mass of synthesised $^{56}$Ni in the explosion can be computed.
Using a fixed density profile allows the abundance distribution in the inner 
layers to be determined. The ratio of the strongest [Fe II] and [Fe III] lines
is a good indicator of the late time ejecta abundances.  This analysis
complements the early time spectra, and yields an abundance distribution
throughout the whole ejecta.

\subsection{Light curve code}

A Montecarlo code is used, with a fixed density profile and derived abundance
distribution, to compute a synthetic bolometric light curve
\citep{Cappellaro97,Mazzali01}. The code computes the emission and propagation
of $\gamma$-rays and positrons as a function of time, and follows their
propagation through the ejecta. When these deposit, optical photons are
produces, whose diffusion is also followed with a Montecarlo scheme.  The
optical opacity is treated with a scheme which is based on the number of
effective lines as a function of abundances \citep{Mazzali01}. This is a good
approximation, as line opacity dominates SNe\,Ia ejecta \citep{Pauldrach96}.

\begin{figure*}
\centering
\includegraphics[scale=0.45]{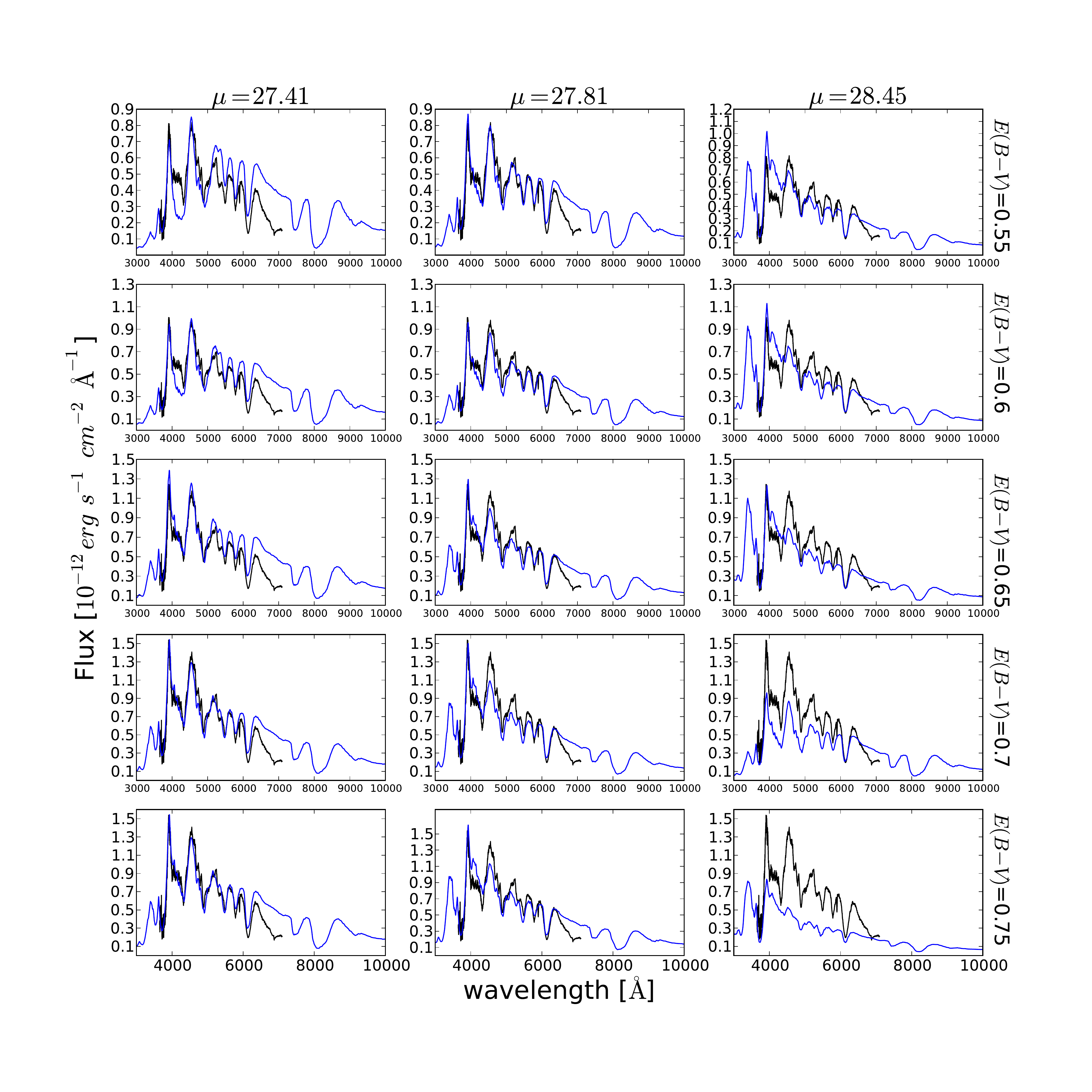}
\caption{One-zone models of SN\,1986G at $B$ band maximum, using a variety of 
distance modulus and extinction values. The observed spectra are in black and 
the modelled in blue. }
\label{fig:distext}
\end{figure*}

\section{Photospheric models}

\subsection{Extinction and distance}

The first step in modelling the photospheric spectra of a SN is to set the
values for host galaxy extinction and distance. Both of these values are very
uncertain for SN\,1986G. A range of values have been published for the distance
modulus of NGC 5128, ranging from 27.41-28.45\,mag (see table
\ref{table:distances3}). A large range of extinction values  have also been
quoted for SN\,1986G, ranging from $E(B-V)_{tot}$=0.6 to 0.9\,mag 
\citep{Nugent05, phillips87}, as well as $E(B-V)_{tot}=0.78\pm0.07\,mag\
(R_{v}=2.4)$ \citep{Taubenberger08}. Therefore, we have taken the distance to
the SN and host galaxy extinction as free parameters. In our models there is
some degeneracy between the distance and the extinction to the SN 
\citep{sasdelli14}. We ran one-zone photospheric phase models on the spectrum at
$B$ band maximum, with fixed abundances, to determine the most likely range of
parameters. Figure \ref{fig:distext} shows some of these one-zone models, for a range of distances from table \ref{table:distances3}. 
 Extinction values range from 0.5 to -0.9\,mag and were
varied by intervals of 0.05\,mag. The range of acceptable values was taken as
$E(B-V)_{tot}$=0.6-0.7\,mag \footnote{0.12\,mag of this is from foreground
galactic extinction \citep{newMWred}} and $\mu=27.41-27.81$\,mag.   For the
modelling in this paper we used $E(B-V)_{tot}=0.65\pm0.5$\,mag and
$\mu=27.61\pm0.4$\,mag, as these values yield the best fits. The extinction is
consistent with that of \citet{Nugent05}.  The distance value is in good agreement
 with the Cepheid distance from \citet{Ferrarese07}, who derive a distance modulus of 27.67$\pm$0.12\,mag. In the modelling process it is most 
important to obtain the correct line ratios, line strengths, ionizations and
velocities, as these are independent of extinction and yield information about
the SN. We find that other values of extinction and distance produce  worse
fits, even if the abundances are varied.

\begin{table}
 \centering
 \begin{minipage}{90mm}
 \caption{Published distances to SN\,1986G.}
  \begin{tabular}{cccc}
  \hline
   Distance modulus ($\mu$)&Method&Source\\
   28.45 $\pm$0.8&Tully Fisher&\footnote{\citet{Richter84}}\\
   27.81$\pm$0.2&TRGB&\footnote{\citet{Soria96}}\\
   27.67$\pm$0.12&Cepheid&\footnote{ \citet{Ferrarese07}}\\
   27.41$\pm$0.04&Brightest Stars&\footnote{\citet{Vaucouleurs80}}\\
\hline
\end{tabular}
\label{table:distances3}
\end{minipage}
\end{table}

\subsection{Density profiles}

Initially, we modelled SN\,1986G using two density profiles, W7 and a
Sub-Ch mass density profile. The W7 density profile results from a fast
deflagration  explosion of a Chandrasekhar  mass C+O WD. The deflagration wave
synthesises 0.5-0.6\,M$_{\sun}$ of $^{56}$Ni in the inner layers of the star,
which is enough to power the light curve of the supernova \citep{w7}. This
explosion has a kinetic energy \KE = $1.3\times 10^{51}$\,erg. The W7 model was
selected as SNe 2004eo, 2003du, 2002bo and 2014J can all be reasonably modelled
with this density profile and variable amounts of \Nifs. The Sub-Ch mass
density profile is an explosion with \KE = $1.2\times10^{51}$\,erg and a total
mass of 1.1\,M$_{\sun}$ \citep{Shigeyama92}. It has a larger density at high
velocities but contains less mass in the inner part of the ejecta compared to
W7. 

\begin{figure*}
\centering
\includegraphics[scale=0.46]{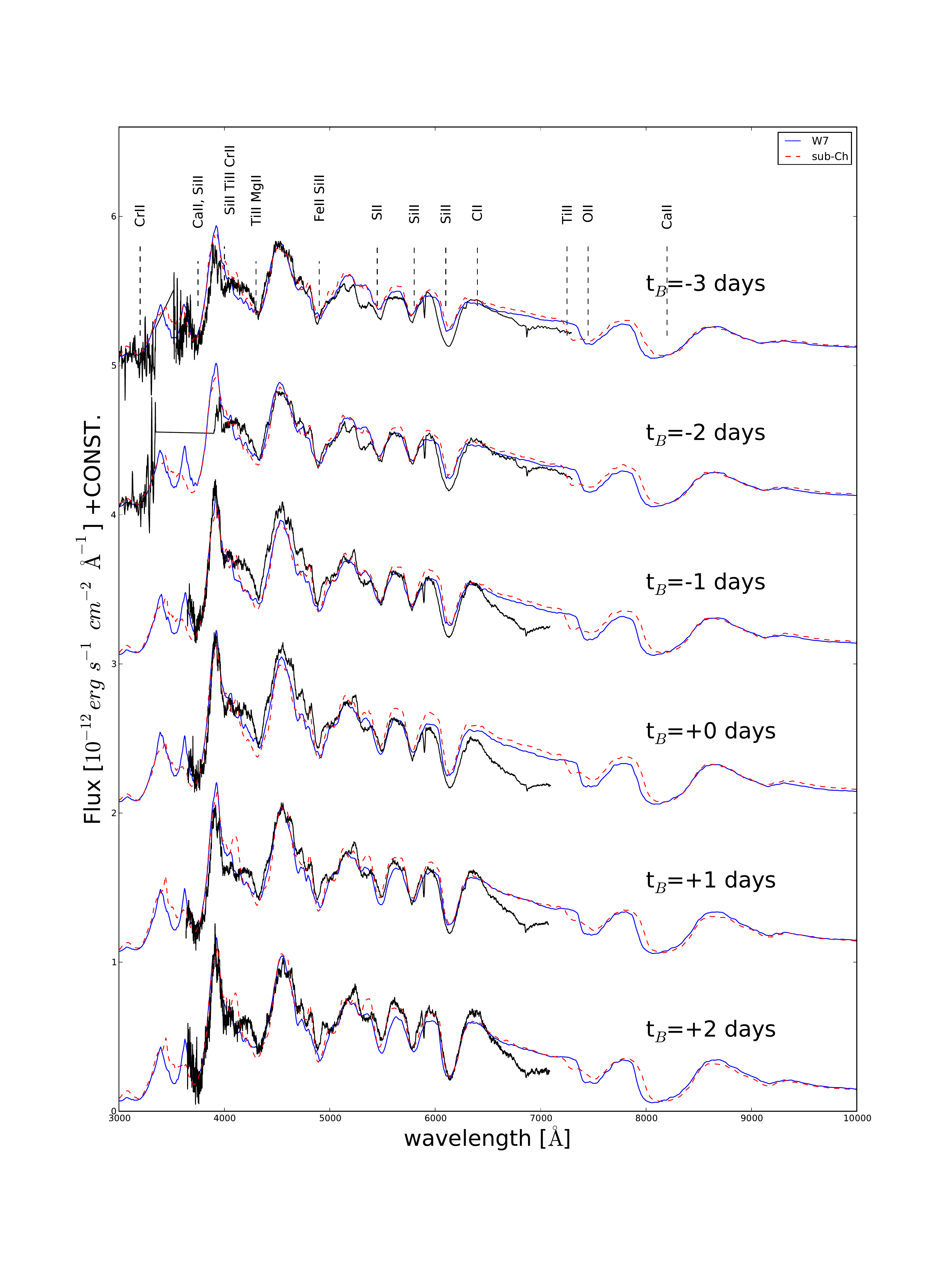}
\caption{The photospheric phase models of SN\,1986G, where the spectra have 
been shifted in flux by a constant for clarity. 
Models for both the W7 (blue solid) and Sub-Ch (red dashed) density 
profiles are shown. The spectra have been corrected for extinction. }
\label{fig:photosphericmodel}
\end{figure*}

\subsection{Photospheric models}

The photospheric phase spectra of SN\,1986G are dominated by lines of
intermediate mass elements (IME), which is consistent with the low $^{56}$Ni
mass and rapid LC decline of the SN, and may indicate a stronger deflagration
phase. This section discusses the input parameters and the properties of the
synthetic spectra at each epoch. Figure \ref{fig:photosphericmodel} presents the
synthetic spectra for the photospheric phase of SN\,1986G.  We adopted for both
the W7 and Sub-Ch mass models a rise time to $B$ band maximum of 18\,days. This
is smaller than for normal SNe\,Ia \citep[\eg][]{Mazzali93}, but is typical for a
SNe\,Ia with a fast LC shape. This is because the ejecta of a low-luminosity SN
have a smaller opacity and a shorter photon diffusion time as they contain less
NSE material  \citep{Mazzali01}. There is very little difference between the two
density profiles within the velocity range sampled by the photospheric models
(7000-10000\,$km\,s^{-1}$). The main differences are in the inner part of the
ejecta, which is sampled by nebular phase modelling. 

Figure \ref{fig:photosphericmodel} shows the spectral evolution of SN\,1986G
between 3 days before and 2 days after $B$ band maximum. The spectra are
dominated by Si II, S II, Fe II, Mg II, Ti II and Cr II lines. Ti II
$\sim$4450\,\AA\, is not typical of normal SNe\,Ia, where Ti is normally doubly
ionised. Most of the lines from Ti III are however in the UV, $<$3500\,\AA.

\subsection{-3 days}

The top spectrum in Figure \ref{fig:photosphericmodel} was observed 3 days
before $B$ band maximum. The synthetic spectrum has a photospheric velocity of 
10000\,km\,s$^{-1}$, and bolometric luminosity of 42.55\,erg\,s$^{-1}$.  The converged temperature of the underlying
blackbody is 8900\,K for the W7 density profile and 9100\,K for the Sub-Ch
density profile. The effective temperature ($T_{eff}$) is 8600\,K for the W7
density profile and 8700\,K  for the Sub-Ch density profile.  For the W7 density
profile the composition is dominated by oxygen (61\% by mass), but it also
requires some unburnt carbon (2\% by mass), IME (Si 26\%, S 6\%, Ca 2\%, Mg
2\%), and traces of heavier metals (Ti+Cr 0.36\% and Fe 0.15\%).  Using the
Sub-Ch density profile, the synthetic spectrum is dominated by O (52\% by mass),
with 2\% unburnt carbon. IME make up a large fraction of the ejecta (Si 25\%, S
15\%, Mg 2\%, Ca 2\%), with the rest consisting of heavier elements (Ti 2\%, Cr
2.5\%, Fe 0.08\%).

\begin{table*}
 \centering

 \caption{Input parameters and calculated converged black body temperatures for models from the W7 and Sub-Ch density profiles.}
  \begin{tabular}{ccccccc}
  \hline
   Epoch&velocity&Bol Lum&Temperature&velocity&Bol Lum&Temperature\\
  $t_{rise}$&$v_{ph}$&L&$T_{BB}$&$v_{ph}$&L&$T_{BB}$\\
  \hline
  days&$km\,s^{-1}$&$logL_{\sun}$&K&$km\,s^{-1}$&$log(L_{\sun})$&K\\
  &W7&W7&W7&Sub-Ch&Sub-Ch&Sub-Ch\\
  \hline
 15&10000&8.960&8900&10000&8.96&9100\\
 16&9400&8.990&9100&9200&9.00&9500\\
 17&8800&9.040&9500&8900&9.040&9600\\
 18&8100&9.060&10000&8500&9.070&9700\\
 19&7800&9.050&9700&7200&9.050&10300\\
 20&7600&9.050&9500&7000&9.050&10100\\
\hline
\end{tabular}
\label{table:W7inputparam}
\end{table*}

\subsection{-2 days}

The second spectrum in Figure \ref{fig:photosphericmodel} was modelled at 16
days after explosion.  There is very little spectral evolution compared to the
previous epoch. Using the W7 density profile a photospheric velocity of 9500 $km
s^{-1}$ is required, as well as a bolometric luminosity of
log$_{10}$L=42.57\,erg\,s$^{-1}$. The $T_{eff}$ at this epoch is 8700\,K. This
spectrum is dominated by oxygen (61\%), while IME make up 38\% of the elements
in this shell (Si 30\%, S 6\%, Ca, 2\%). The Ti, Cr and stable Fe abundances
have all increased (Ti+Cr 1.1\%, Fe 0.2\%), relative to the higher velocity
spectrum. This is required as the Ti $\sim$4450\,\AA\ feature progressively gets
stronger over time. The IUE spectrum has been used to set the flux level in the
UV and to constrain the Cr abundance. 

The Sub-Ch density profile produces slightly different results. The effective
temperature at this epoch is 9200\,K, and the photospheric velocity is
9200\,km\,s$^{-1}$. This layer of the ejecta requires no carbon, and is
dominated by oxygen (50\%). In this shell the S abundance drops to 5\%, whereas
the Si abundance increases to 35\%. This is unusual for a SN\,Ia as one would
expect S to propagate to slightly lower velocities than Si as a consequence of
the nucleosynthesis reaction chain described by \citet{w7}. However, the S II
feature at 5640\AA\ is fit almost perfectly by the model at this epoch, and 
increasing the S abundance would worsen the fit.  It should be noted that in the
abundance distributions of SNe 2004eo and 2011fe the S abundance does not
propagate as deep as that of Si \citep{Mazzali08,Mazzali15}.  Using the Sub-Ch
density profile the abundances of Ti+Cr (8\%) and Fe (1.3\%) are higher than those needed for the W7 model. 
The Ti +Cr abundance is required to be this high to fit the blue ward edge of the
4200\AA\ feature. Without this high abundance there is no absorption at 4100\AA\ . 

\subsection{-1 days}

The third spectrum in Figure \ref{fig:photosphericmodel} was taken at -1 day
relative to $B$ band maximum.  This spectrum was modelled as 17 days after
explosion. The synthetic spectrum produced using the W7 density profile is a
very good fit to the observed one. The feature at $\sim$\,4900\AA\ has resolved
into separate Fe II lines, and the W7 density profile fits all of these lines at
the correct velocities. The photospheric velocity at this epoch is
8900\,kms$^{-1}$, with a bolometric luminosity of 42.63\,erg\,s$^{-1}$  and a $T_{eff}$ of 9200\,K. This shell is
still dominated by oxygen, however at this epoch the abundance of oxygen is
starting to decrease. The oxygen fraction is 50\% by mass. The Si abundance has
increased to 35\%, and that of Ti+Cr to 1.5\%. 

The Sub-Ch model has $v_{ph} = 8900$\,kms$^{-1}$ and a bolometric luminosity of
log$_{10}L$=42.60\,erg\,s$^{-1}$.  At this epoch oxygen still dominates at 44\%,
but IMEs also significantly contribute to the spectrum (Si 37\%, S 5\%, Ca 2\%).
Relative to the previous epoch the Ti+Cr abundance has decreased to 6.5\%, but
the Fe abundance has increased to 5\%.  The effective temperature at this epoch
is 9400\,K.

\subsection{+0 days}

IMEs begin to dominate the ejecta at this epoch, which is at $B$ band and
bolometric maximum. For the W7 model, the Si and S abundances are 65\% and 13\%,
respectively. This spectrum was computed with a photospheric velocity of
8100\,kms$^{-1}$ and a bolometric luminosity of log$_{10}L$=42.65\,erg\,s$^{-1}$. The effective temperature at this epoch is 9700\,K.
The combined Ti+Cr abundance has increased to 1.8\%. 

The Sub-Ch model for this spectrum has a log$_{10}L$=42.66\,erg\,s$^{-1}$, a $v_{ph} =8500$\,kms$^{-1}$ and an effective temperature of
9600\,K, the photospheric velocity has hardly changed compared to the previous
epoch. IMEs also begin to dominate the Sub-Ch model (Si 40\%, S 5\%, Ca 8\%).

 \subsection{+1 days}

This spectrum was observed at +1 day relative to $B$ band maximum. The
photospheric velocity (7800\,kms$^{-1}$) has only slightly decreased (by
300\,kms$^{-1}$ ) compared to the previous epoch.  The model at the epoch has a luminosity of  log$_{10}L$=42.64\,erg\,s$^{-1}$. Because of the small
change in photospheric velocity from the previous epoch there is practically no
spectral evolution. Therefore the abundances have not changed significantly. The
effective temperature of this shell is 9600\,K. 

The Sub-Ch model at this epoch contains no S near the photosphere, although
there is still 34\% of oxygen left in the ejecta. This may indicate that the
Sub-Ch density is not a good solution and cannot explain the mechanism and
progenitor of SN\,1986G.  There is an increase in Fe abundance at this epoch 
(7\%), and the Ti+Cr abundance is 4\%. The effective temperature of the Sub-Ch
model is 10200\,K.  The model at the epoch has a luminosity of  log$_{10}L$=42.64\,erg\,s$^{-1}$, and a photospheric velocity of 7200\,kms$^{-1}$.

\subsection{+2days}

The last W7 photospheric model for the spectrum at +2 days (20 days from
explosion). It is shown in Figure \ref{fig:photosphericmodel}. It has a
$v_{ph}$=7600\,km\,s$^{-1}$,  $T_{eff}$=9300\,K and
log$_{10}L$=42.62\,erg\,s$^{-1}$. This shell has 75\% Si and 10\% S. The Ca
abundance is set at 0.3\%.

The final Sub-Ch model still has a large mass of oxygen (30\%), but this epoch
is dominated by Si (56\%). The Ti and Cr abundances have stayed constant, but
the Fe abundance has increased to 8\%. The effective temperature at this epoch
is 10100\,K.  There are some doubly ionized species in the spectrum at this
time, Fe III and Cr III in the features at $\sim$\,5000\AA\ and $\sim$3700\AA\
respectively. These doubly ionized species correspond to the effective
temperature increase, compared to the +1\,days model. The W7 model has fewer
doubly ionized species. 

At every epoch in the Sub-Ch model there is a deep Ti II absorption at
$\sim$\,7300\AA\. This feature is not seen in the W7 models, nor in the observed
spectrum, and it is produced because the Sub-Ch model has more mass at high
velocities.  It is another indication the Sub-Ch density profile is a poor
solution compared to the W7 model. However, apart form this feature, it could be
argued that the Sub-Ch density profile produces better fits. The full input
parameters for the photospheric models can be found in Table
\ref{table:W7inputparam}. The photospheric velocity of SN\,1986G is lower than
what would be expected for a  normal SNe\,Ia.  The photospheric phase models
probe layers above 7000\,km\,s$^{-1}$ in velocity. The bolometric luminosity of
the models peak at +0 days compared to $B$ band maximum, which is expected from
a SN with a steep light curve.

\section{Nebular phase models}

A single nebular spectrum of SN\,1986G is available, which was published by
\citet{Cristiani92} and was obtained $\sim 256$ days after maximum. The spectrum
shows the usual nebular SN\,Ia features, strong \FeII\ and \FeIII\ lines in the
blue and a mix of \CaII\ and \FeII\ lines in the red. It is significantly affected
by reddening, to the point of showing self-absorption in the NaID line as well
as a narrow \Ha\ absorption in the middle of a weak \FeII\ emission complex. As a
result of the large amount of reddening, lines in the blue appear to be
suppressed.

We have completed the abundance tomography experiment modelling the nebular
spectrum using two different density distributions: the original W7 model and
the Sub-Ch model discussed above. 

We used the nebular spectrum code which has been used in the past to model both
SNe\,Ia \citep[e.g.][]{Mazzali15} and SNe\,Ib/c \citep[e.g.][where a more
detailed description of the code can be found]{mazzali07}. The code computes the
diffusion and deposition of gamma-rays and positrons produced by the radioactive
decay of $^{56}$Ni and $^{56}$Co in the SN ejecta using gray opacities in a
montecarlo scheme. The ensuing collisional heating of the SN ejecta is then
balanced by cooling via line emission, following \citet{Axelrod80}. The balance
between radiation and gas properties is computed in non-local thermodynamic
equilibrium (NLTE). The abundances in the ejecta were adjusted to optimise the
fit, except at the velocities where abundances were determined from early-time
spectral modelling. These are regions with velocities above 7000\,kms$^{-1}$,
which are not sampled by the nebular emission lines. A simple one-zone fit to
the spectrum yields a model line velocity of 5600\,kms$^{-1}$, which is also in
line with the fast evolution of the light curve of SN\,1986G
\citep{mazzali98,mazzali07}.

\begin{figure}
\centering
\includegraphics[scale=0.45]{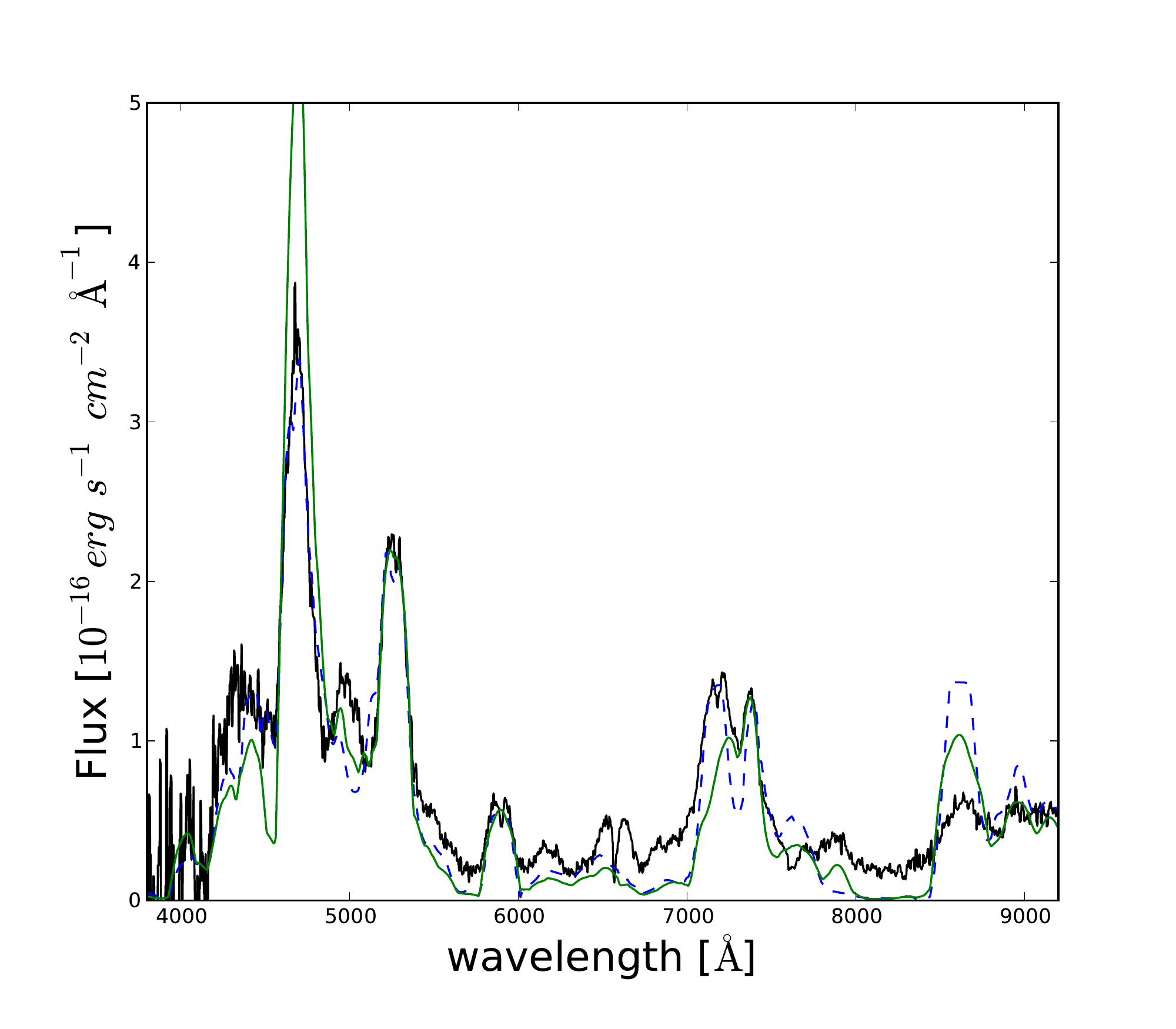}
\caption{The nebular phase models of SN\,1986G.
 The blue spectrum was obtained
using a W7 density profile and the green spectrum using the Sub-Ch profile. 
The black line is the observed spectrum. The spectra have been corrected for extinction.}
\label{fig:nebcomp}
\end{figure}

The two best-fit models are shown in Figure \ref{fig:nebcomp}. Although the
models are overall quite similar in the kind of emission lines that are
predicted, there are significant differences which can help us discriminate
among them. The individual models are discussed in turn.

The W7 model (dashed/blue line) produces a good match in particular to the blue
part of the spectrum. A $^{56}$Ni mass of 0.13 M$_{\sun}$ is required. Because
the inner region in W7 is quite dense, material other than $^{56}$Ni must be
used to fill it. Stable Fe-group elements are produced in the centre of
Chandrasekhar-mass explosions, and we find that a total mass of about
0.11M$_{\sun}$ is required. This is dominated by stable Fe. Stable Fe-group
material is important as is acts only as a coolant, so its presence contributes
to balancing the heating from radioactive decay and to keeping the ionization
ratio (in particular \FeIII/\FeII) close to the observed value. The mass
included within a velocity of 5600\,kms$^{-1}$ in W7 is $\sim 0.4 M_{\sun}$, and
NSE material does not reach this value. The remaining mass is attributed to
intermediate-mass elements, in particular Si, which leads to a strong predicted
emission in the NIR, near 1.6$\mu$\,m. The line near 5900\,\AA\ is predominantly
NaID.  The strong emission near 7300\,\AA\ is well reproduced as a combination
of \FeII\ and \CaII\, with a minor contribution from \NiII\ . The narrowness of the
emission lines leads to the observed split between the two features. In fact, a
hint of a split can be seen even in the emission feature near 5200\,\AA, and
this is due just to lines separating out, as in, e.g., SN\,2003hv
\citep{Mazzali11}, not to double peaks caused by the morphology of the
explosion. Other \FeII\ lines are not well reproduced, though. In particular, the
feature near 6600\,\AA\ is too weak in the model, while the broad blend near
9000\,\AA\ is too strong. These shortcomings are seen in most of our SN\,Ia
models, but to a much lesser degree \citep[e.g.][]{Mazzali15}. They may
partially depend on uncertain collisional rates, and they indicate that the
$^{56}$Ni mass may be slightly overestimated. On the other hand, the extreme
deviation of the redder lines in SN\,1986G may also indicate problems with the
density structure \citep[see e.g.][]{Mazzali11}, or issues with the red part of
the observed spectrum. Note that in a well-observed SN\,Ia like SN\,2011fe the
ratio of the Ca-Fe emission near 7300\,\AA\ and that of the Fe complex near
9500\,\AA\ is $\sim 1$. In the spectrum of SN\,1986G, after correcting for
reddening, the ratio is $\sim 2$, but our model has a ratio of $\sim 1$. 
Unfortunately no other nebular spectrum is available of either SN\,1986G or of
SNe that closely resemble it, so we cannot verify this. 

The Sub-Ch model (green line), compared to the W7 one, has several
problems. Primarily, the ionization of Fe is too high. This is shown by the
excessive strength of the emission near 4700\,\AA\ when the model matches that
at 5200\,\AA. Note that the model displayed is reddened, which depresses the
bluer feature significantly. This behaviour was seen in a peculiar SN like
2003hv \citep{Mazzali11}, and was interpreted there to indicate low densities at
low velocities, which is exactly what a Sub-Ch model predicts, and a
consequent lack of stable NSE material, leading to insufficient cooling and
hence recombination. The \FeII\ lines in the red are actually depressed relative
to the line at 5200\,\AA. The model contains only $\sim 1$\,M$_{\sun}$. Most of
the missing mass is stable NSE material, the mass of which has gone down to
0.1\,M$_{\sun}$, mostly stable Fe at intermediate velocities. Intermediate-mass
elements just above the $^{56}$Ni zone again lead to strong Si emission in the
NIR. The incorrect ratio of the optical Fe lines is a strong argument against
this model, which we do not favour.

\section{Abundance tomography}

\begin{figure*}
\centering
\includegraphics[scale=0.5]{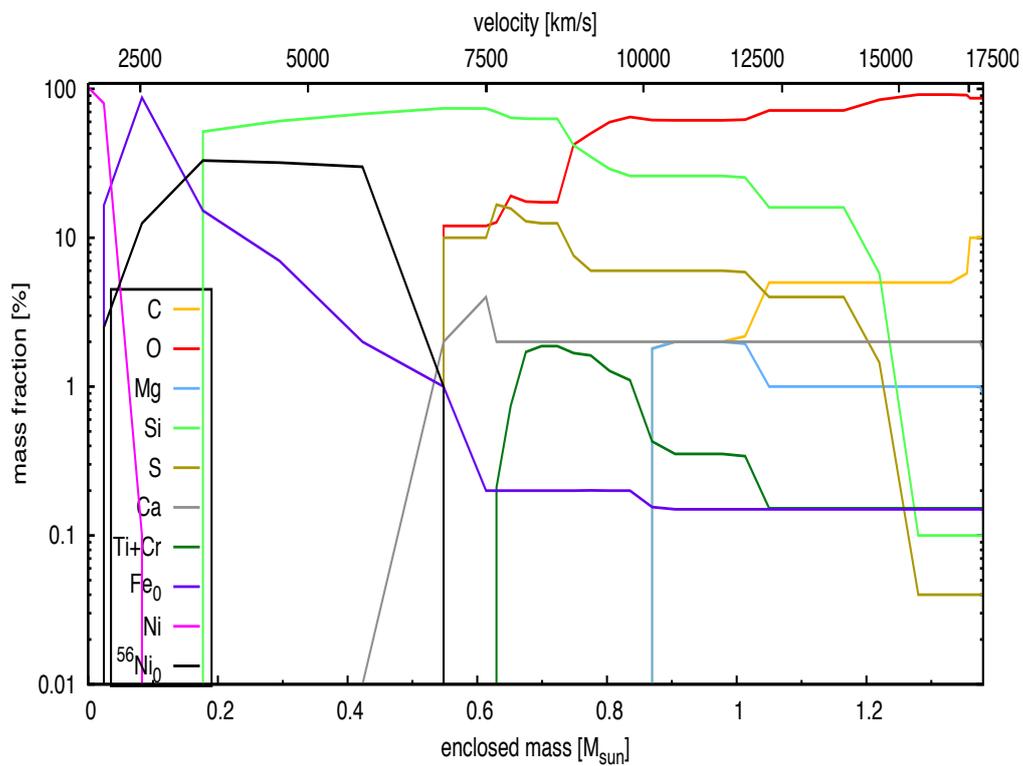}
\caption{The final abundance distribution of SN\,1986G obtained with the W7 
density profile. }
\label{fig:W7abund}
\end{figure*}

\begin{figure*}
\centering
\includegraphics[scale=0.5]{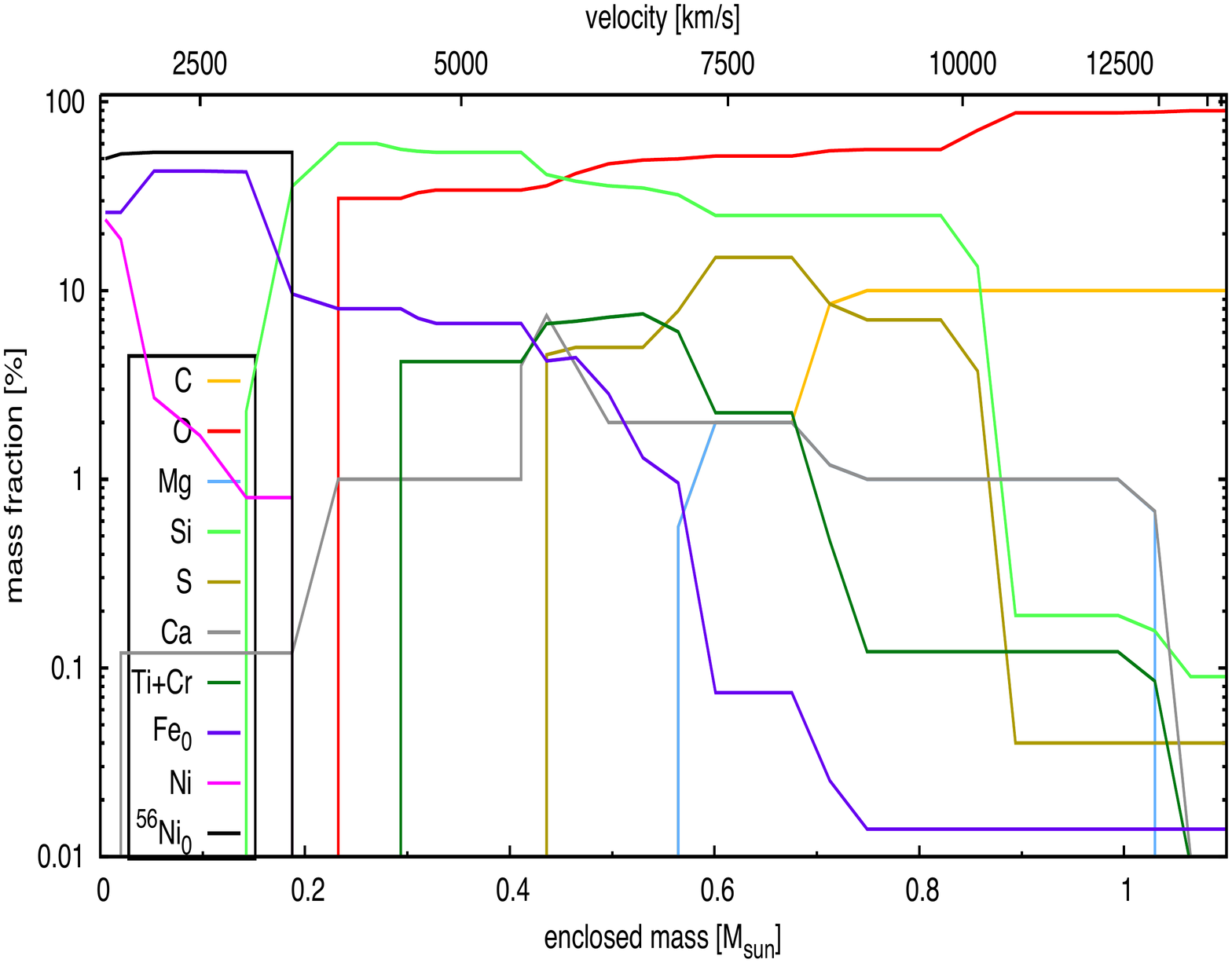}
\caption{The final abundance distribution of SN\,1986G obtained with the 
Sub-Ch density profile. }
\label{fig:subabund}
\end{figure*}

Figures \ref{fig:W7abund} and \ref{fig:subabund} show the abundance
distributions of SN\,1986G as a function of velocity and enclosed mass. At
velocities below 7000\,kms$^{-1}$ the nebular modelling determines the
distribution, and at velocities above 7000\,kms$^{-1}$ the photospheric models
determine the distribution.  Above $\sim$12000\,kms$^{-1}$ certain aspects of
the distribution can be inferred, but this is where the results are most
uncertain because early data are not available. However, incorrect abundances in
the outer layers can make it impossible to produce a good synthetic spectrum in
the inner layers, so we can assume that the description of the outer layers is
not completely unreasonable. 

Figure \ref{fig:W7abund} is the abundance distribution produced using the W7
density profile. An upper limit to the abundance of carbon is set to a mass of 0.02\,M$_{\sun}$. All
of this progenitor carbon is at high velocities, $>$10000\,kms$^{-1}$.  The outer layers of the ejecta are dominated by oxygen, which
is seen throughout the whole of the photospheric phase, but the part of the
observed spectra where we would expect to see oxygen, $\sim$7500\,\AA, was not
observed.  It is reasonable to infer a high oxygen abundance in SN\,1986G,
because of its similarities with 91bg-like SNe, which show a strong oxygen
feature.Furthermore, the fact that we detect carbon in the spectra is a good indication that 
oxygen will be present between layers where the carbon and IME dominate.  
The final masses from the W7 abundance distribution can be found in Table
\ref{table:intabun}. Masses have been quoted to the second decimal place, but
does not mean accuracy to this degree. It is required as there are elements
which do not have a mass greater than 0.1\,M$_{\sun}$. IMEs dominate the
abundance distribution of the SN at intermediate velocities.  The S distribution
follows the Si distribution at velocities down to $\sim$7000\,kms$^{-1}$, with a
lower ratio than traditionally would be expected. The typical ratio of Si to S
is 3 to 1. The S abundance drops to 0 at $v \sim$6000\,kms$^{-1}$.  The enclosed
mass range in which Si dominates is from 0.75\,M$_{\sun}$ to 0.19\,M$_{\sun}$.
This is very different from SN\,1991T, which sits at the opposite end of the
`Phillips Relation', where Si dominates only in the enclosed mass range of
$\sim$1.05\,M$_{\sun}$ to 1.07\,M$_{\sun}$ \citep{sasdelli14}. The large amount
of Si in SN\,1986G corresponds to a small amount of $^{56}$Ni being produced in
the explosion \citep{mazzali07}. This agrees with SN\,1986G having a low luminosity
and partial burning.  There is no evidence for mixing of $^{56}$Ni to high
velocities for SN\,1986G.  A $^{56}$Ni mass of 0.13\,M$_{\sun}$ is obtained from
the abundance distribution. Most \Nifs\ is located in the denser inner layers.
At velocities between $1000$\,km\,s$^{-1}$ and $\sim3500$\,km\,s$^{-1}$ Fe and
$^{56}$Ni dominate the ejecta. The innermost layers of the SN\,1986G W7 density
profile model are dominated by stable Ni, and $^{56}$Ni is further out in the
ejecta, between an enclosed mass of $\sim$0.1 and $\sim$0.55$\,M_{\sun}$. This
is consistent with the burning that a fast deflagration such as that of a
W7-like model predicts. The Ti+Cr abundances peak in the velocity range 
8000-9000\,kms$^{-1}$. Their combined integrated mass is of the order of
0.01\,M$_{\sun}$. 

The abundance distribution produced using the Sub-Ch density profile  is
dramatically different from that of the W7 density profile, as shown in Figure
\ref{fig:subabund}. The total C mass is 0.04\,M$_{\sun}$. There is evidence for
carbon down to an enclosed mass of $\sim$0.7\,M$_{\sun}$.  The outer layers
consist almost entirely of oxygen. The oxygen zone is large compared to a normal
SN Ia.  Oxygen dominates down to 6000\,km\,s$^{-1}$, and is present down to 
velocities as low as 4000\,kms$^{-1}$ with abundances of the order of 20\%. 
Between 6000\,km\,s$^{-1}$ and 4000\,km\,s$^{-1}$ IMEs dominate, and NSE
material dominates the inner layers of the ejecta. The total amount of $^{56}$Ni
in the ejecta of SN\,1986G, assuming a Sub-Ch density profile, is
0.11\,M$_{\sun}$.  The abundance distributions produced using the Sub-Ch model
implies that this explosion scenario is not a valid one for SN\,1986G.  Oxygen
probes to deeper layers of the ejecta than sulphur, which is in direct conflict
with nucleosynthesis calculations and expected results. The zone where IMEs dominate has an
unusually small velocity range of 500\,kms$^{-1}$. Overall, the abundance yields
and distribution from the Sub-Ch model do not agree with nucleosynthesis and
explosion models.

\begin{table}
 \centering
 \caption{Integrated abundances from the full abundance tomography modelling of 
SN\,1986G. The errors on the masses are $\pm$25\%, except for \Nifs\ which has an error of $\pm$10\%.}
  \begin{tabular}{ccc}
  \hline
   Element&W7&Sub-Ch\\
   &M$_{\sun}$&M$_{\sun}$\\
  \hline
C&0.02&0.04\\
O&0.49&0.51\\
Mg&$<$0.01& $<$0.01\\
Si&0.50&0.26\\
S&0.05&0.04\\
Ca&0.02&0.01\\
Ti+Cr&0.01&0.02\\
Fe&0.11&0.09\\
$^{56}$Ni&0.13&0.11\\
Ni&0.04&0.01\\
\hline
$M_{tot}$&1.38&1.11\\
\hline
\end{tabular}

\label{table:intabun}
\end{table}

\section{A consistent, reduced-energy model}

\subsection{Energy estimates}

With the integrated masses obtained form the abundance tomography modelling, the
kinetic energy of the explosion can be derived, using the formula  

\begin{equation}
E_{k}=[1.56(^{56}Ni)+1.74(stableNSE)+1.24M(IME)-E_{BE}]10^{51}\,erg
\label{eq:KE}
\end{equation}

\citep{Woosley07}. Where $E_{BE}$ is the binding energy of the progenitor white
dwarf, $^{56}Ni$ is the $^{56}Ni$ mass in the ejecta, stableNSE is the stableNSE 
mass is the ejecta and M(IME) is the total IME mass in the ejecta.
 We find that the \KE\ of the explosion 
is $0.8 \times 10 ^{51}$\,erg, using the masses derived by the W7 density profile model
 \footnote{For a Chandrasekhar mass WD $E_{BE}
=0.46 \times 10 ^{51}$\,erg.}. The \KE\ calculated from the abundances obtained
using the W7 density profile is smaller than the value for the standard W7
explosion, which has \KE\ $= 1.3\times$10$^{51}$\,erg. Therefore, there is a
discrepancy between the input density profile and the \KE\ derived from the
nucleosynthesis.  This discrepancy is an issue if one wants to solve the
progenitor scenario and explosion mechanism of SN\,1986G.  For the Sub-Ch
density model $E_{BE} =0.21 \times10^{51}$\,erg \citep{Yoon05}, which equates to
a \KE\ of $0.55\times10^{51}$\,erg (the \KE\ of the Sub-Ch explosion model that
we used is $1.2\times10^{51}$\,erg), so again the model has a larger \KE\ than
what we infer from the nucleosynthesis.

We therefore attempted to develop a density profile that both fits the spectra
and is consistent with the nucleosynthesis. Given what we discussed above, this
has to be a low energy density profile. In this section we build and test such a
profile. We did this by scaling the W7 density profile to a lower \KE\ using
equations \ref{eq:rho} and \ref{eq:energy}.

\begin{equation}
\rho^\prime=\rho_{W7} \Big( \frac{E^\prime}{E_{W7}}\Big)^{-\frac{3}{2}}\cdot\Big( \frac{M^\prime}{M_{W7}}\Big)^{\frac{5}{2}}
\label{eq:rho}
\end{equation}

\begin{equation}
v^\prime=v_{W7} \Big( \frac{E^\prime}{E_{W7}}\Big)^{\frac{1}{2}}\cdot\Big( \frac{M^\prime}{M_{W7}}\Big)^{-\frac{1}{2}}
\label{eq:energy}
\end{equation}

\begin{figure}
\centering
\includegraphics[scale=0.35]{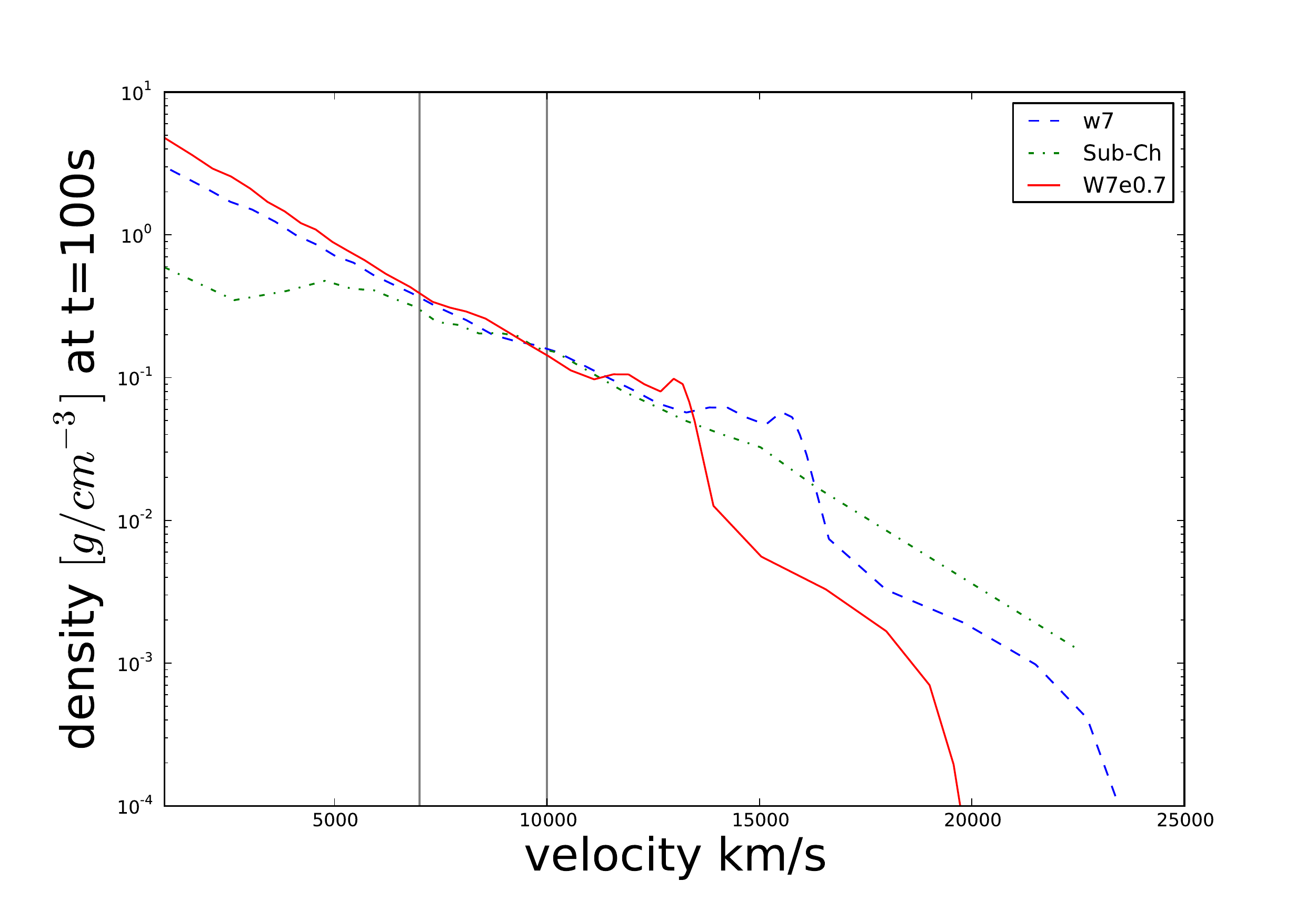}
\caption{The W7, Sub-Ch and W7e0.7 density profiles as a function of velocity 
at $t$=100\,s after explosion. The vertical grey lines show the range in values 
$v_{ph}$ can take, for the -3\,d to +2\,d models. }
\label{fig:denistyprofile}
\end{figure}

In these equations $\rho^\prime$ is the new density, $\rho_{W7}$ is the W7
density, $E^\prime$ is the energy of the new density profile, $E_{W7}$ is the
energy of the W7 density profile, $M^\prime$ is the mass of the new density
profile and $M_{W7}$ is the mass of the W7 density  profile, $v_{W7}$ is the
velocity of W7 the density profile, and $v^\prime$ is the velocity of the new
profile. 

The density profile was scaled keeping the mass at the Chandrasekhar mass and
scaling the energy to 70\% of that of the W7 density profile to match the \KE\
inferred from the nucleosynthesis, and for convenience it is called W7e0.7. This
rescalilng has been shown to work well for SN\,2005bl. However, this SN was only
modelled in the photospheric phase \citep{Hachinger09}, as late-time spectra
were not available. Therefore, placing the mass in the inner layers may not have
been the perfect solution \citep[see][]{Mazzali11}. In contrast, SN\,1986G has
good nebular data, so it is possible to tell if this increase in central density
can be a realistic solution. 

Figure \ref{fig:denistyprofile} shows the W7, Sub-Ch and W7e0.7 density profiles
at $t$=100\,s after the explosion. Compared to W7, the Sub-Ch density profile
has more mass at higher velocities and a lower central density, whereas the
W7e0.7 model has a higher central density and less material at high velocities,
and hence a smaller \KE. To decrease the energy in the W7 model, the mass which
was removed from the high velocities was redistributed to the inner part of the
ejecta.

\subsection{Photospheric-epoch models} 

\begin{figure}
\centering
\includegraphics[scale=0.3]{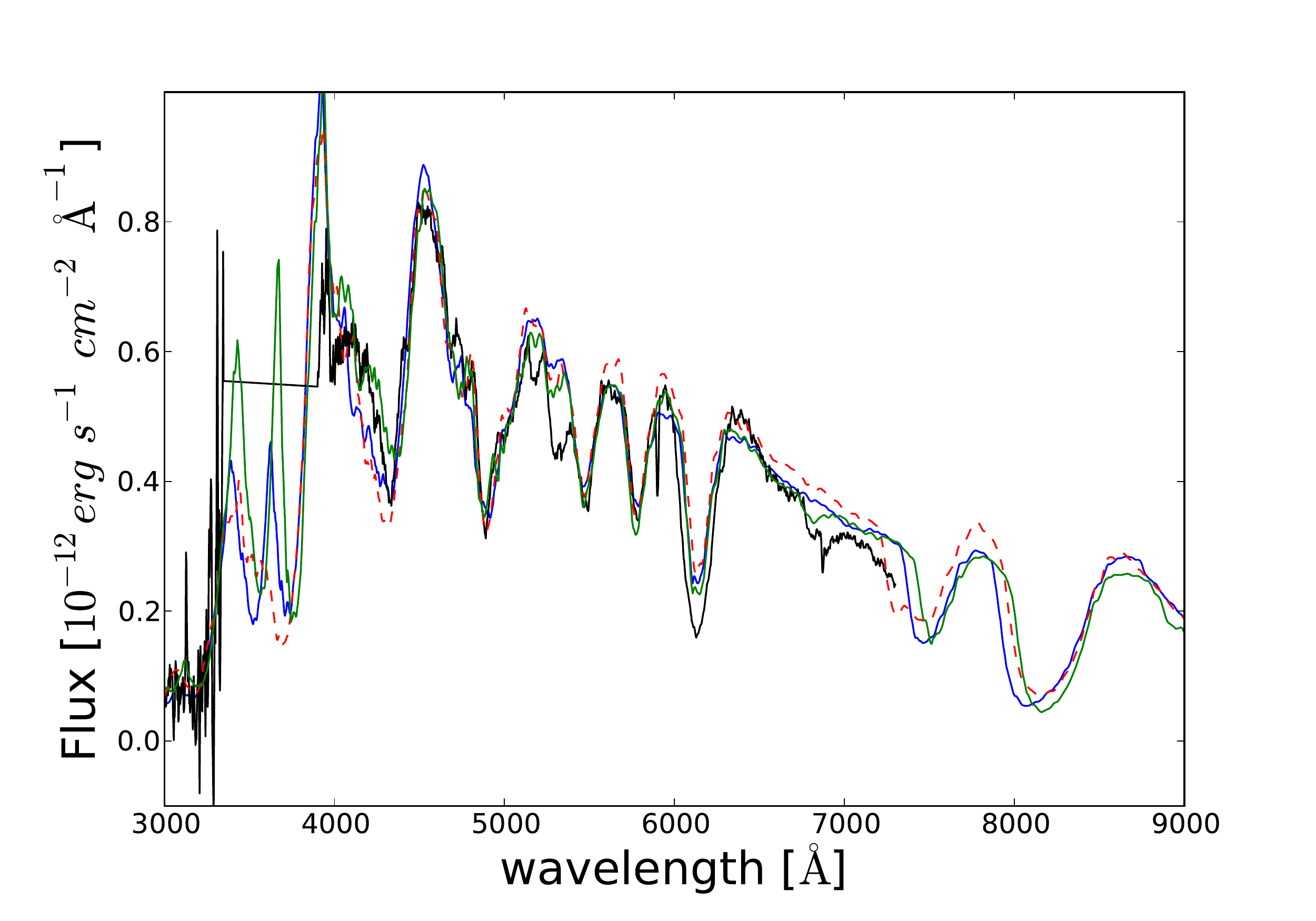}
\caption{One-zone models at -3 days relative to $B$ band maximum. The synthetic spectra were produced using the W7 (blue), W7e0.7 (green) and Sub-Ch (red dashed) density profiles. The spectra have been corrected for extinction.}
\label{fig:lowonezone}
\end{figure}

Figure \ref{fig:lowonezone} shows the synthetic spectra produced with all three density profiles at -3\,days relative to $B$ band maximum. The green line is the W7e0.7 model. This model fits the data (black line) very well, the \SiII\ line ratio is correct, and the sulphur and Fe features are fit almost perfectly. The fact that the W7e0.7 density profile produces a good fit is reason for the full abundance tomography analysis to be carried out using the W7e0.7 density profile, with the aim of getting a consistent model.

\begin{figure*}
\centering
\includegraphics[scale=0.5]{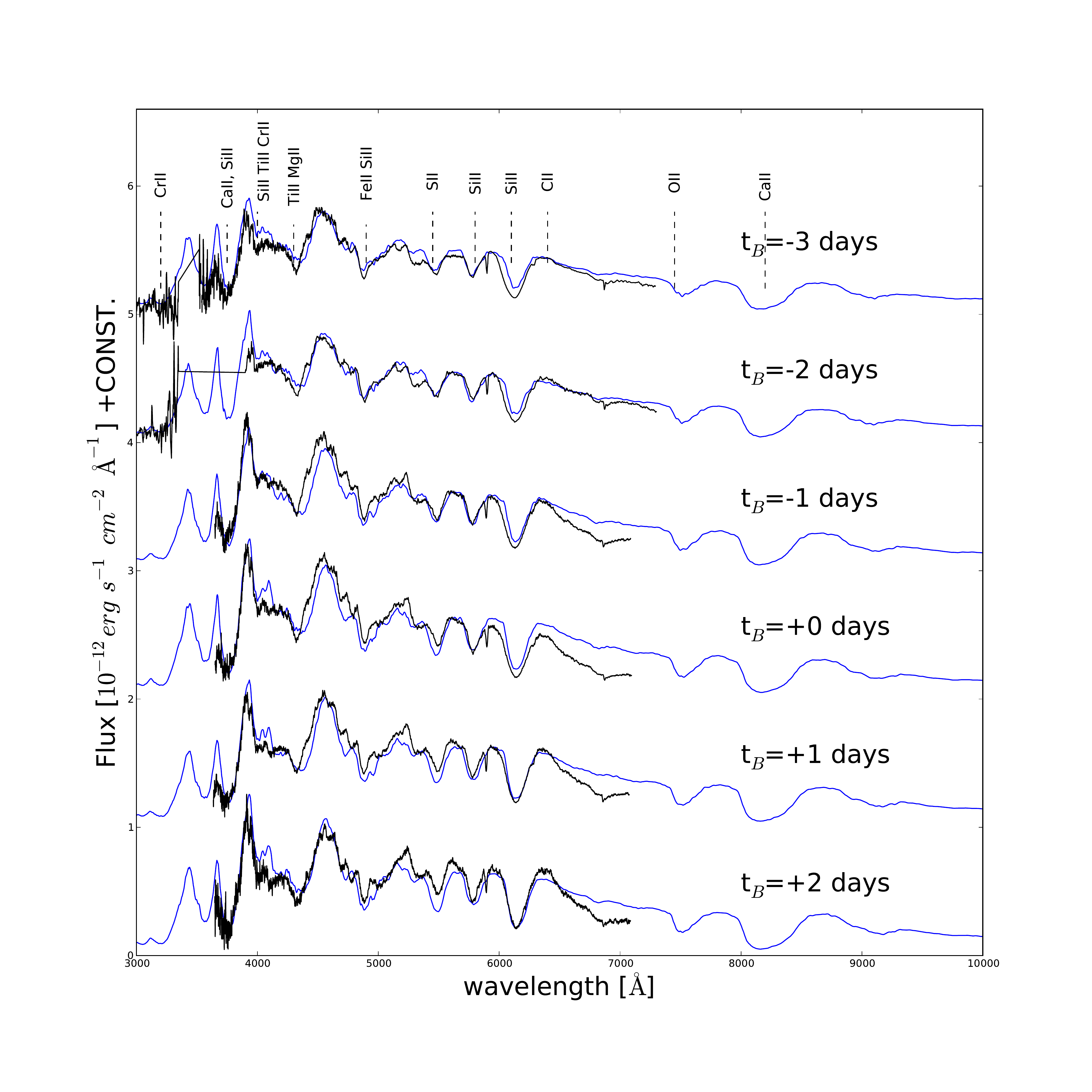}
\caption{The photospheric phase models of SN\,1986G, calculated using the 
W7e0.7 density profile. The blue line are the models and the black line the 
observed spectra. The spectra have been corrected for extinction.}
\label{fig:lowenergyphot}
\end{figure*}
 
The full photospheric models obtained with the new, W7e0.7, density structure are presented in
Figure \ref{fig:lowenergyphot}. The synthetic spectra are better
than the Sub-Ch and W7 equivalent. The models were remade using the same input
parameters as those for the W7 model in Table \ref{table:W7inputparam}. The only elements
which changed in abundance, in velocity space, were Ti+Cr, with their mass fraction increasing to
$\sim3\%$ at 8000 km\,s$^{-1}$. In the W7e0.7 spectra, the ratio of the
strongest Si II features (6355 and 5970\,\AA) has improved and the Ti
$4450\,\AA$ feature fits significantly better. These improvements could be
caused by the bump in the density profile moving inwards in velocity space. This
causes there to be an increase in density closer to the photosphere. This bump
is still in the oxygen zone, but there is also an increase in the Si abundance. 
The improvements are also due to a lack of material at high velocities. It is
apparent that a steeper density than the standard W7 model is required above
12500 kms$^{-1}$. 

\subsection{Carbon}

\begin{figure}
\centering
\includegraphics[scale=0.4]{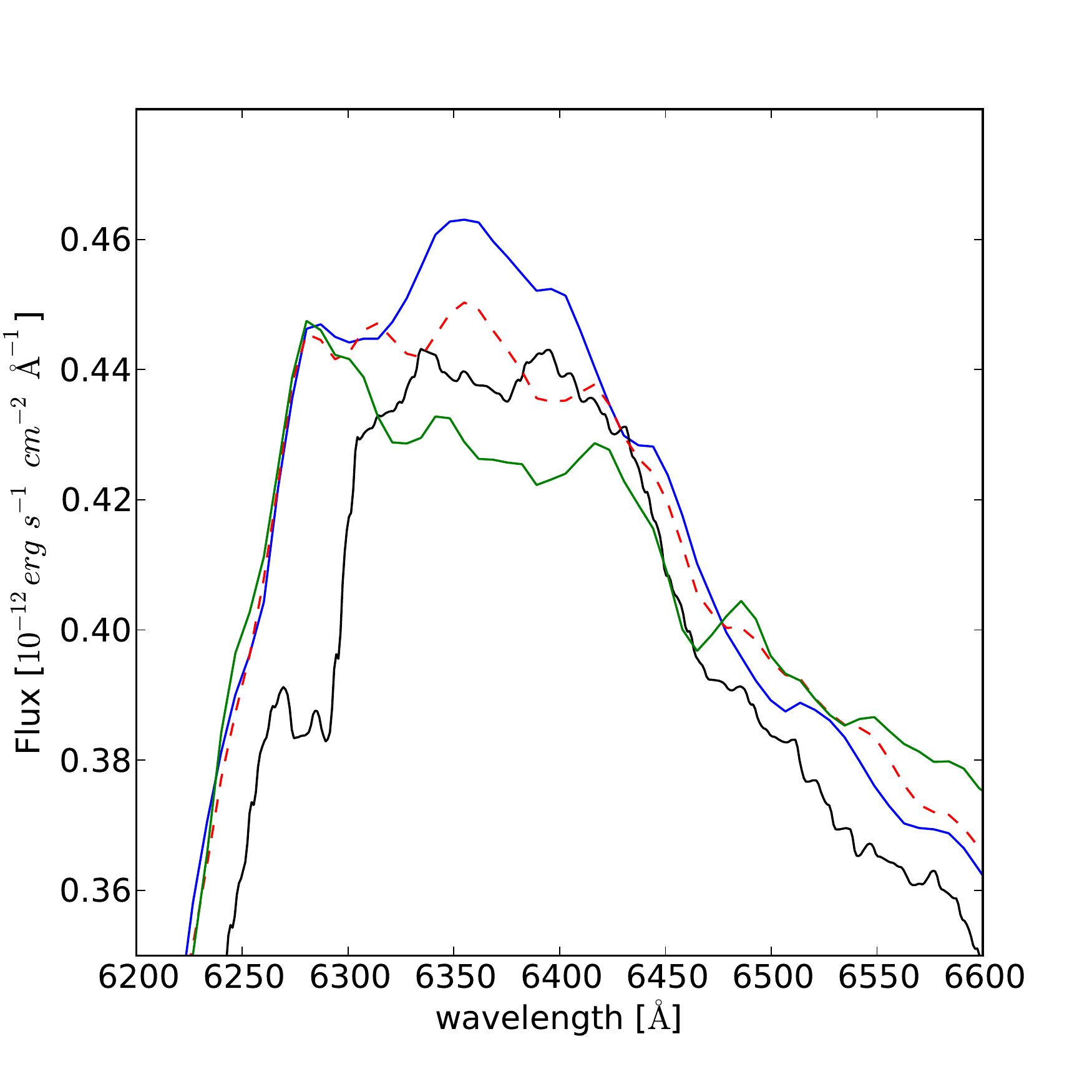}
\caption{Synthetic spectrum at -3 days produced using the W7e0.7 density 
profile to constrain an upper limit on carbon. The blue line is the model with 0\% 
of C, the red dashed line with the 2\% C and the green line with 10\% C at the photosphere. The black line is the observed 
spectra.}
\label{fig:W7carb}
\end{figure}

From Figure  \ref{fig:photosphericmodel} it is apparent at -3 and -2 days that there is some C II in the spectra, at $\sim6350$\AA. 
As there are no observed spectra before -3 days it is not possible to put an exact constraint on the amount of unburnt carbon in the
 ejecta, therefore in the next part of the analysis we attempt to constrain an upper limit on the C abundance.  
 The synthetic spectra produced using the W7e0.7 density profile has 3 synthetic shells 
based above the -3 days, 10000\,kms$^{-1}$, outer shell. The purpose of having these synthetic layers is to produce a stratified 
abundance distribution. If the distribution is not stratified the model spectra will produce a poor fit. Figure \ref{fig:W7carb} shows the 
-3 day spectrum over a wavelength range of 6200-6600\,\AA, where unburnt carbon would be seen in the ejecta. In order to produce
 the best fit of this C II feature the W7e0.7 model requires a carbon abundance of 13\% at 23000\,kms$^{-1}$, 10\% at 17550\,kms$^{-1}$,
 5\% at 12000\,kms$^{-1}$ and 2\% at 10000\,kms$^{-1}$. Increasing the 12000\,kms$^{-1}$ layer to an abundance of 12\% produces 
the green line in Figure  \ref{fig:W7carb}, and having zero carbon in the ejecta produces the blue line in Figure \ref{fig:W7carb}, both of
 which yield a worse fit. When determining the best fit it is important to only examine the spectra between 6300-6450\AA\, as this is where the C II $\lambda$\,6578 and $\lambda$\,6582 lines will be seen, blue-ward of this is Si II absorption and red-ward is continuum. The integrated carbon mass abundance from the best fit using the W7e0.7 density profile is 0.02\,M$_{\sun}$.

\subsection{Nebular-epoch model} 

\begin{figure}
\centering
\includegraphics[scale=0.45]{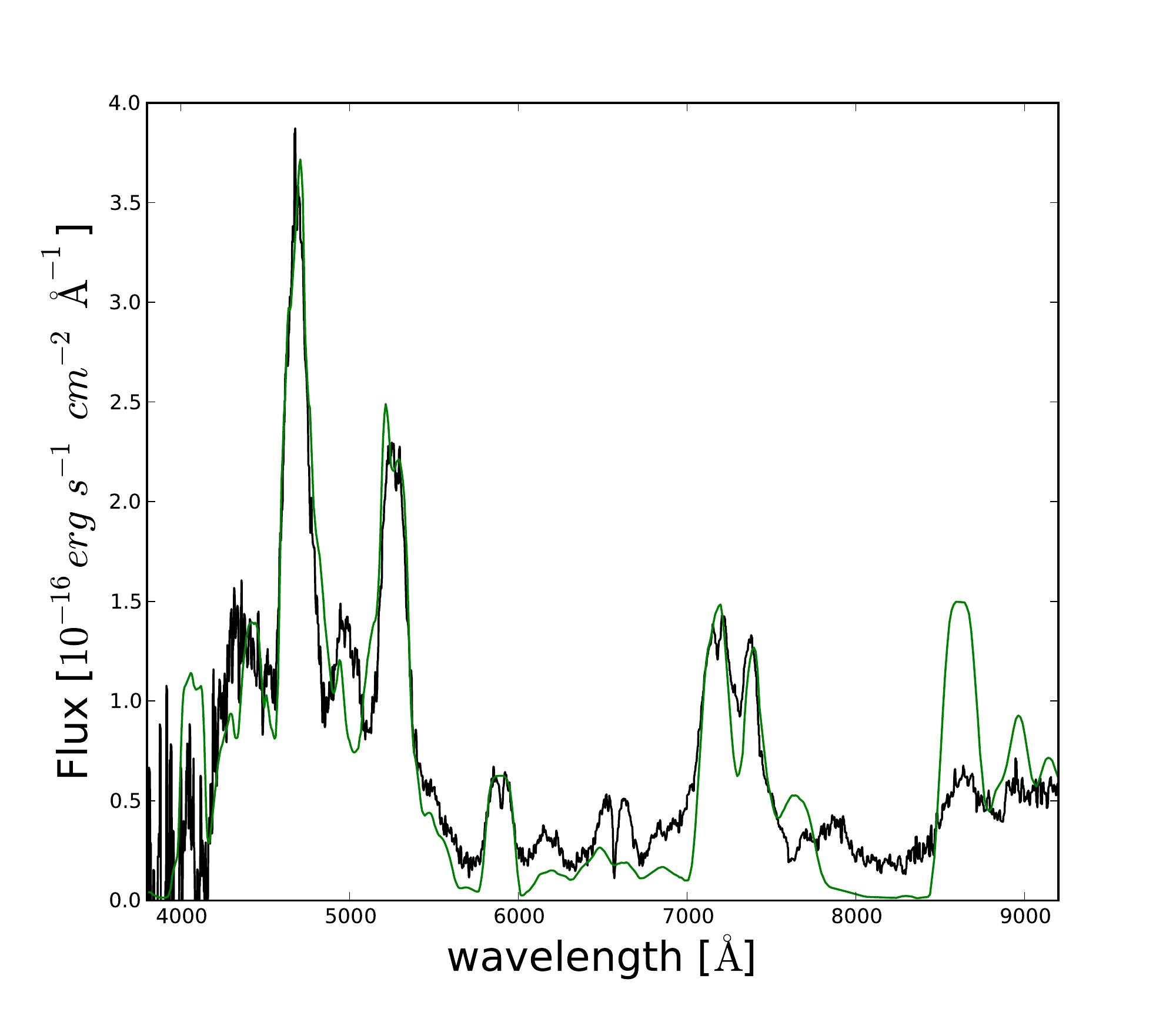}
\caption{The nebular phase models of SN\,1986G. The green line is the model calculated using the W7e0.7 density profile.
 The spectra have been corrected for extinction.}
\label{fig:lowenergyneb}
\end{figure}
 
A nebular phase model was produced using the W7e0.7 density. It is shown as a
green line in Fig. \ref{fig:lowenergyneb}. This density profile has more mass at
low velocity ($\sim$ 0.50 M$_{\sun}$\ within 5600\,kms$^{-1}$ as opposed to
$\sim 0.40$ for W7). The $^{56}$Ni mass is now $\sim$ 0.14 M$_{\sun}$, but the
higher density at low velocities results in a larger mass of stable NSE
material, $\sim$ 0.21 $M_{\sun}$, most of which is stable Fe. This is in line
with \citet{mazzali07}.  The additional cooling provided by the stable NSE
material leads to the correct reproduction of the \FeII/\FeIII\ line ratio in
the blue, but it does not solve the problem of the excessive \FeII\ emission in
the red, where the synthetic spectrum predicts excessively strong lines.  As we
argued above, there may be a problem with our model, but also with the redder
part of the observed spectrum itself.  Nebular-epoch observations of
SN\,1986G-like SNe are encouraged in order to assess the reality of the red
flux.

\subsection{Abundance stratification} 

\begin{figure*}
\centering
\includegraphics[scale=0.5]{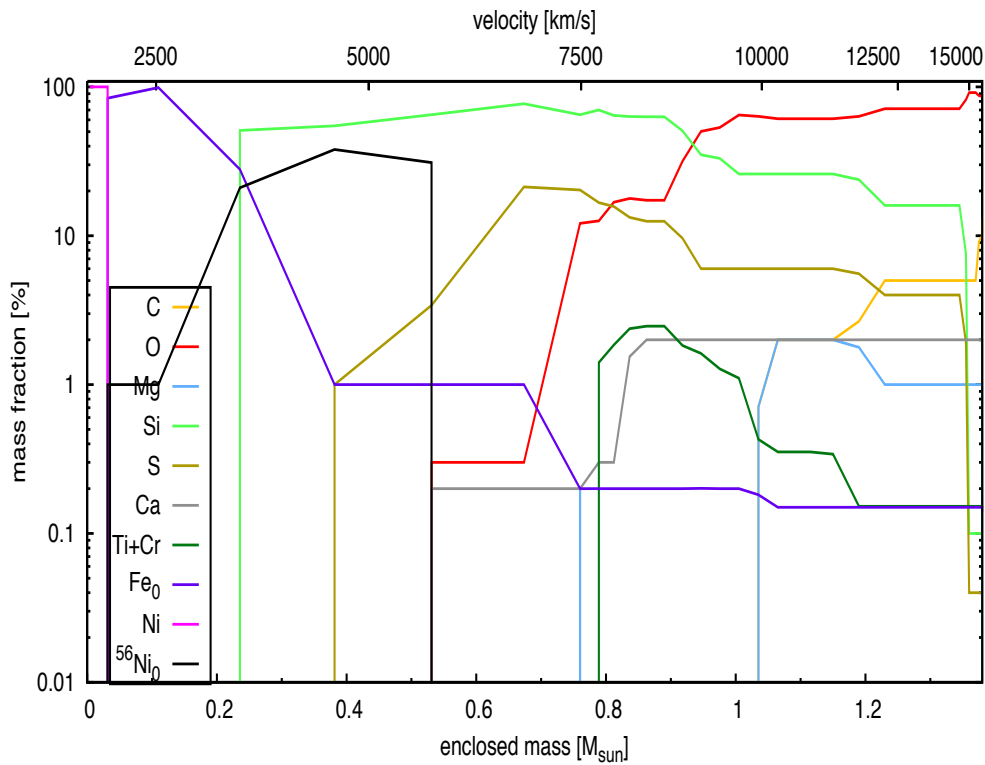}
\caption{The abundance distribution of the ejecta of SN\,1986G obtained 
using the W7e0.7 density profile.}
\label{fig:lowenergyabund}
\end{figure*}

The abundance stratification of the W7e0.7 model can be seen in Figure
\ref{fig:lowenergyabund}. The oxygen layer dominates the ejecta down to
velocities of $\sim9000$\,km\,s$^{-1}$. IMEs dominate the bulk of the ejecta in
at velocities between $\sim9000$\,km\,s$^{-1}$ and $\sim3500$\,km\,s$^{-1}$.
Below $\sim3500$\,km\,s$^{-1}$ NSE material dominates, with the ejecta
consisting entirely of stable Ni in the innermost $1000$\,km\,s$^{-1}$.
$^{56}$Ni is more evenly distributed than in the W7 model. The $^{56}$Ni
abundance peaks at 40\% in the velocity range $4000 - 6000\,km\,s^{-1}$.  
It is also seen in the ejecta down to a velocity of $1000$\,kms$^{-1}$. In the
models produced using the W7 and Sub-Ch density profiles oxygen probes to deeper
layers of the ejecta than sulphur, which is in direct conflict with SNe\,Ia
nucleosynthesis calculations and explosion models \citep{Iwamoto99}. However, in
the W7e0.7 model sulphur is required down to a velocity of
$\sim$4000\,kms$^{-1}$, and oxygen is only seen down to velocities of
$\sim6000$\,kms$^{-1}$, so the problem no longer exists. 

The abundance distribution from the W7e0.7 model produces physically sensible
results, as well as good fits between the observed and synthetic spectra. The
integrated masses obtained using this density profile can be found in Table
\ref{table:intabun70}. The Ni mass is 0.14\,M$_{\sun}$, with IMEs making up
0.69\,M$_{\sun}$ and 0.34\,M$_{\sun}$ of unburnt material. Using equation
\ref{eq:KE} the kinetic energy of the ejecta can be calculated. The \KE\ 
calculated using this method is $0.97\times10 ^{51}$\,erg, which is consistent
with the energy of the W7e0.7 density profile ($0.9\times10 ^{51}$\,erg).

\begin{table}
 \centering
 \caption{Integrated abundances the the models which used the W7e0.7.
 The errors on the masses are $\pm$25\%, except for \Nifs\ which has an error of $\pm$10\%.}
  \begin{tabular}{cc}
  \hline
   Element&W7 E70\%\\
   &M$_{\sun}$\\
  \hline
C&0.01\\
O&0.33\\
Mg&$<$0.01\\
Si&0.58\\
S&0.09\\
Ca&0.01\\
Ti+Cr&0.01\\
Fe&0.18\\
$^{56}$Ni&0.14\\
Ni&0.02\\
\hline
$M_{tot}$&1.38\\
\hline
\end{tabular}

\label{table:intabun70}
\end{table}

\section{Bolometric Light curves}

\subsection{Observed light curve}

We have constructed a bolometric light curve of SN\,1986G in the range
3000-24000\,\AA, covered by our optical and NIR photometry, and used the
well-monitored SN\,Ia 2004eo as a proxy.

As a first approach, we dereddened the optical ($UVB$) and NIR ($JHK$) light
curves of SN\,1986G using $E(B-V)_{tot}$ = 0.65 and the extinction curve of
\citet{CCMred}. We then splined these points with a time resolution of 1 day and
constructed daily spectral energy distributions in the above wavelength interval
using the flux zeropoints of \citet{Fukugita95}.  For each epoch, we integrated
the flux between the U and K bands after interpolating the flux between the
central wavelengths of the filters, and added at the blue and red boundaries of
the interval the fluxes obtained by extrapolating the spectrum with a flat 
power-law to 3000 and 24000\,\AA, respectively.  We then resampled the final
bolometric LC to the epochs of the actual V-band observations.

The lack of significant coverage in $R$ and $I$ bands (only a couple of points
are available in each of these bands) represents a serious drawback, because the
interpolation between $V$ and $J$ bands overestimates the $R$ and $I$ flux. 
Therefore, the final bolometric LC must be considered as an upper limit to the
"real" one (which we will call `$UL_{bol}86G$').  On the other hand, if we only
integrate separately the flux in $UBV$ and in $JHK$ filters over the
3000-6000\,\AA\ and 10000-24000\,\AA\ ranges, respectively, and sum these two
broad-band fluxes, we neglect completely the flux contribution in $R$ and $I$
bands, so that the resulting bolometric LC must be considered as a lower limit
of the true one (`$LL_{bol}86G$'). Clearly, the real bolometric LC must be
somewhere between $UL_{bol}86G$ and $LL_{bol}86G$.

To evaluate the necessary bolometric corrections in the most rigorous way
possible and thus to obtain a reliable estimate of the bol LC of SN\,1986G, we
have resorted to the SN Ia 2004eo, which had good coverage in all bands between
$U$ and $K$ \citep{Pastorello07,Mazzali08}, as an analogue of SN\,1986G.  Since
$U$ and $JHK$ bands coverage for SN\,2004eo is only available after maximum, we
reconstructed the flux in these bands before maximum assuming that the $U-B$,
$J-I$, $H-I$ and $K-I$ colours before maximum are constant and equal to the
values they have at maximum (we have verified a posteriori that this has a
negligible effect with respect to simply ignoring the $UJHK$ band before maximum
in the computation of the bolometric flux).  The $UBVRIJHK$ light curves of
SN\,2004eo, dereddened with E$(B-V)$ = 0.109, were splined with a time
resolution of 1 day, and spectral energy distributions were constructed and
integrated over the range 3000-24000\,\AA, as for SN\,1986G. This is our
reference bolometric light curve, which we will call our `template' bolometric
LC.

We have then constructed two more bolometric light
curves for SN2004eo in the same wavelength intervals as
ULbol86G and LLbol86G above,  and ignoring the R and
I bands:  first we  interpolated the spectral flux between V and J
- we will call this ULbol04eo -  and then we summed the spectral
flux in the UBV filters (3000-6000\AA) and in the JHK filters
(10000-24000\AA).
We will call this latter bolometric LC LLbol04eo.
   Bolometric corrections for SN\,2004eo with respect to the
`template' were computed by simply differencing the  template and its upper
($UL_{bol}04eo$) and lower ($LL_{bol}04eo$) limits.  Then, these differences
were applied, with their signs, to the limits obtained for SN\,1986G,
$UL_{bol}86G$ and $LL_{bol}86G$, so that 2 `corrected' bolometric LCs for
SN\,1986G were obtained.  What we consider the `real'  bolometric LC of
SN\,1986G was obtained by averaging these 2 `corrected' bolometric light curves.

\subsection{Synthetic light curve}

In the spirit of abundance tomography, we now combine the density distributions
that we used and the abundances derived through synthetic spectra fitting and
compute a bolometric light curve, which we compare to the one constructed based
on the observed SN photometry. We use a montecarlo code \citep{Cappellaro97}
which computes the emission of gamma-rays and positrons following the
radioactive decay of $^{56}$Ni and $^{56}$Co and their deposition in the ejecta.
The $^{56}$Ni distribution is derived from the spectral fitting. Gamma ray
deposition is computed with a grey opacity of 0.027\,cm$^2$\,g$^{-1}$, and for
positrons a grey opacity 7\,cm$^2$\,g$^{-1}$ is used.

After deposition of the gamma-ray and positron energy, the resulting energy is
assumed to be thermalised and energy packets representing photons are then
followed as they propagate through the SN ejecta. The opacity they encounter is
assumed to be dominated by line opacity and is parametrised according to the
relative abundances at different depths according to the number of active lines
as in \citet{mazzali2000}: \\
$\kappa = 0.25 Ab(Fe-gp) + 0.025 \times (1 - Ab(Fe-gp))$

\begin{figure}
\centering
\includegraphics[scale=0.4]{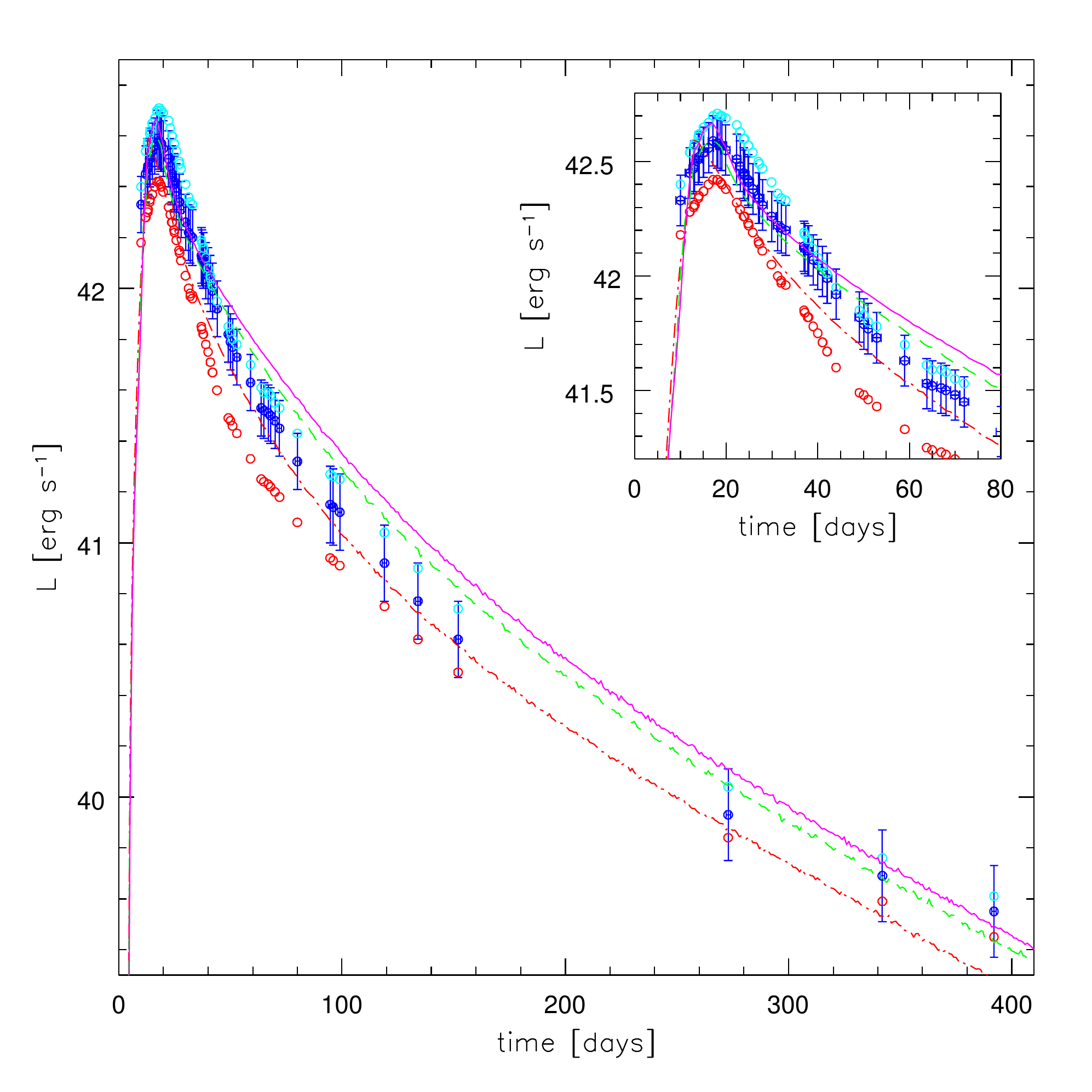}
\caption{The observed (blue dots) bolometric light curve with the upper (cyan 
markers) and lower (red markers) limits. As well as the modelled 
LC using the derived abundance distribution from the W7 (green), Sub-Ch (red) 
and W7e0.7 (purple) density profiles. }
\label{fig:boldLC}
\end{figure}

The results for the different models are shown in Figure \ref{fig:boldLC} along
with the observationally derived bolometric light curve. At peak, the Sub-Ch
model (red/dot-dashed line) reaches maximum within 12-13 days, which appears to
be a short time compared to the observed light curve. The early declining part
of the light curve also happens too early in this model, but then at epochs
between 60 and 300 days this model follows the data reasonably well, although it
underestimates the luminosity because of the low gamma ray depositions. At late
times, the light curve matches the data at 275 days, and is again too low later
on, but at these epochs the observed light curve declines unusually slowly. 

Both Chandrasekhar-mass models, on the other hand, reach maximum after about
17-18 days, which is in line with the observed light curve and the results from
spectroscopic modelling. These models also follow the early light curve decline
quite well, but then are more luminous than the observed light curve by almost a
factor of 2 after about day 60. At late times they finally rejoin the observed
light curve, but they have a different slope. Looking in more detail, the
low-energy model (purple/continuous line) is somewhat more luminous than the W7
model (green/dashed line) after day 60, because of the enhanced gamma-ray
deposition resulting from the higher densities in the more slowly expanding
ejecta. This difference is however quite small, such that neither model can be
favoured over the other. The indications for a low energy from spectral fitting
are therefore not disproved by the light curve model.

The real issue is the behaviour of the light curve, first after 60 days and then
at late times. The steep decline after day 60 is not in line with the behaviour
of other SNe\,Ia, \citep[e.g. SN\,2004eo]{mazzali07}, which suggests that the IR
contribution is not known. The range of possible corrections does in fact allow
a much higher luminosity. At the latest times the decline is quite slow,
possibly indicating incorrect background subtraction. The one point when spectra
are available still suffers from the lack of IR information. Again, observations
of a modern 1986G-like SN\,Ia would be needed to improve both the data coverage
and the modelling results. 

Yet, our results indicate that a low-energy Chandrasekhar-mass model is favoured
over an undermassive model, which may come somewhat as a surprise given that
SN\,1986G showed quite a rapidly evolving light curve. We argue that the rapid
light curve evolution is the result of the low opacity, as a
consequence of the low NSE elements abundance ($\sim 0.35 \Msun$).

\section{Conclusions}

A full theoretical spectral analysis of SN\,1986G has been performed, using the
`abundance tomography' approach.  SN\,1986G bridges the gap between a normal and
subluminous SN\,Ia. It is found that the ejecta of SN\,1986G have a low
temperature and are dominated by singly ionised IMEs. Three density profiles
were tested, a standard Chandrasekhar mass delayed detonation (W7) density
profile, a low energy Chandrasekhar mass delayed detonation (W7e0.7) density
profile and a Sub-Chandrasekhar mass detonation density profile (Sub-Ch).

The Sub-Ch models produce good photospheric phase synthetic spectra,
 but this density profile can be ruled out owing to the unrealistic
abundance distribution and poor synthetic nebular spectra. Furthermore, the W7
density profile is found to not be the ideal solution to explain the explosion
of SN\,1986G. There is no consistency between the \KE\ implied by the density
distribution ($1.3\times10^{51}$\,erg) and that calculated using the abundance
distribution ($0.8\times10^{51}$\,erg).  

It was found that a low energy W7 profile (W7e0.7) produced the best results.
The fact that the W7e0.7 model yields a sensible abundance distribution and a
consistent model demonstrates that any deviation from this mass should be
minor.  The final integrated masses of the various elements that were obtained
from the W7e0.7 model are: O+C=0.34\,M$_{\sun}$, IME=0.69\,M$_{\sun}$, stable
NSE=0.21\,M$_{\sun}$ and  $^{56}$Ni=0.14\,M$_{\sun}$. These abundances produce a
\KE\ of $0.97\times10^{51}$\,erg, which is consistent with the energy of the 
W7e0.7 density profile ($0.9\times10^{51}$\,erg).

In conclusion, SN\,1986G is a low-energy Chandrasekhar mass explosion of a C+O WD, which produced a
small amount of $^{56}$Ni and a large amount of IMEs. The spectra of SN\,1986G
show signs of progenitor \CII\ as late as  -3 days from $B$ band maximum.
Although SN\,1986G was a rapidly-declining SN\,Ia, we find no reason for it to
deviate from a Chandrasekhar mass explosion. SN\,1986G may be classified as the
extreme end of the normal population of SNe\,Ia, rather than part of the
sub-luminous population. It is even more peculiar in properties than SN\,2004eo.
However, like SN\,2004eo its low luminosity can be interpreted as the explosion
having a strong deflagration phase, which unbinds the expansion of the star,
reducing the density and therefore the effectiveness of the successive
supersonic burning phase. This led to the production of a large fraction of IMEs
\citep{Mazzali08}. 

These findings raise the possibility that only SNe Ia with very large decline
rates deviate from a Chandrasekhar mass. The uncertainties in the analysis lie
in the value of host galaxy extinction and the gaps in spectral coverage.
However, SN\,1986G is the only SN\,Ia with properties falling in the gap between
normal and subluminous SNe\,Ia that also has a good time-series of spectra. Thus
our analysis of SN\,1986G explores an important region in parameter space in the
overall SNe\,Ia picture.

\section*{Acknowledgments}
Some of the data presented in this paper were obtained from the Mikulski 
Archive for Space Telescopes (MAST). STScI is operated by the Association of 
Universities for Research in Astronomy, Inc., under NASA contract NAS5-26555. 
Support for MAST for non-HST data is provided by the NASA Office of Space 
Science via grant NNX09AF08G and by other grants and contracts. We have used 
data from the NASA / IPAC Extragalactic Database (NED, 
http://nedwww.ipac.caltech.edu, operated by the Jet Propulsion Laboratory, C
alifornia Institute of Technology, under contract with the National Aeronautics 
and Space Administration). For data handling, we have made use of various 
software (as mentioned in the text) including \textsc{iraf}. \textsc{iraf}  
Image Reduction and Analysis Facility (http://iraf.noao.edu) is an astronomical 
data reduction software distributed by the National Optical Astronomy 
Observatory (NOAO, operated by AURA, Inc., under contract with the National 
Science Foundation).

\bibliography{bibtex}
\end{document}